\documentclass[aps,prb,reprint,superscriptaddress,amsmath]{revtex4-2}
\usepackage[utf8]{inputenc}
\usepackage{graphicx}
\usepackage{enumerate}
\usepackage{refcount}
\usepackage{amsthm}
\usepackage{amssymb}
\usepackage{dsfont}
\usepackage{hyperref}
\usepackage{physics}
\usepackage{bbm}
\usepackage{dcolumn}
\usepackage{longtable}
\usepackage{tablefootnote}
\usepackage[usenames,dvipsnames]{color}
\usepackage[normalem]{ulem}
\usepackage{tikz}
\def\checkmark{\tikz\fill[scale=0.4](0,.35) -- (.25,0) -- (1,.7) -- (.25,.15) -- cycle;} 

\hypersetup{pdfstartview=FitH,pdfpagemode=UseOutlines,colorlinks,
citecolor=blue,linkcolor=blue,urlcolor=blue}

\newcommand{\MnPSe}{MnPSe$_3$}

\begin{document}

\title{Controlling Topology through Targeted Composite Symmetry Manipulation in Magnetic Systems}

\author{Ilyoun Na}
    \email{ilyoun1214@berkeley.edu}
    \affiliation{Department of Physics, University of California, Berkeley, California 94720, USA}
    \affiliation{Materials Sciences Division, Lawrence Berkeley National Laboratory, Berkeley, California 94720, USA}
    \affiliation{Molecular Foundry, Lawrence Berkeley National Laboratory, Berkeley, CA 94720, USA}
 
\author{Marc Vila}
    \email{marcvila@berkeley.edu}
    \affiliation{Department of Physics, University of California, Berkeley, California 94720, USA}
    \affiliation{Materials Sciences Division, Lawrence Berkeley National Laboratory, Berkeley, California 94720, USA}

\author{Sin\'{e}ad M. Griffin}
    \email{SGriffin@lbl.gov}
    \affiliation{Materials Sciences Division, Lawrence Berkeley National Laboratory, Berkeley, California 94720, USA}
    \affiliation{Molecular Foundry, Lawrence Berkeley National Laboratory, Berkeley, CA 94720, USA}

\date{\today}   

\begin{abstract}
The possibility of selecting magnetic space groups by orienting the magnetization direction or tuning magnetic orders offers a vast playground for engineering symmetry protected topological phases in magnetic materials. In this work, we study how selective tuning of symmetry and magnetism can influence and control the resulting topology in a 2D magnetic system, and illustrate such procedure in the ferromagnetic monolayer \MnPSe. Density functional theory calculations reveals a symmetry-protected accidental semimetalic (SM) phase for out-of-plane magnetization which becomes an insulator when the magnetization is tilted in-plane, reaching band gap values close to $100$ meV. We identify an order-two composite antiunitary symmetry and threefold rotational symmetry that induce the band crossing and classify the possible topological phases using symmetry analysis, which we support with tight-binding and $\mathbf{k}\cdot\mathbf{p}$ models. Breaking of inversion symmetry opens a gap in the SM phase, giving rise to a Chern insulator. We demonstrate this explicitly in the isostructural Janus compound Mn$_2$P$_2$S$_3$Se$_3$, which naturally exhibits Rashba spin-orbit coupling that breaks inversion symmetry. Our results map out the phase space of topological properties of ferromagnetic transition metal phosphorus trichalcogenides and demonstrate the potential of the magnetization-dependent metal-to-insulator transition as a spin switch in integrated two-dimensional electronics.
\end{abstract}

\maketitle

\section{Introduction}
There has been remarkable progress in understanding the interplay between symmetry and topology in matter, resulting in a systematic classification of topological states of matter based on the symmetries they possess. A subset of these -- magnetic topological materials -- can host a plethora of topological phases ranging from Weyl and Dirac semimetals, to Chern and axionic insulators, and have been identified in both experiments and theory \cite{Frey2020, Xu2020, Elcoro2021, Bernevig2022}.

The recent experimental discovery of two-dimensional (2D) magnetic materials~\cite{Gong2017, Huang2017} -- once thought to be precluded by the Mermin-Wagner theorem -- has opened up many new horizons for both fundamental quantum materials research and their applications. For instance, such low-dimensional magnetic materials have great potential as ultracompact magnetic memories and spin-torque devices~\cite{wang2013low, Kurebayashi2022, Yang2022}, even possessing hard ferromagnetism down a monolayer\cite{Husremović2022}. Moreover, these discoveries have in turn been followed by the identification of topological phases in 2D magnetic systems including the realization of 2D topological magnetic textures~\cite{Wu2020, Ding2020}, topological superconductivity~\cite{Li2019, Kezilebieke2020}, topological semimetals (SM)~\cite{You2019, Niu2019, Xu2022}, time-reversal-broken topological insulators~\cite{Niu2020, Wang2020} and Chern insulators~\cite{Mishra2018, Sugita2018, Kang2023}.

The presence of topological phases in 2D materials is particularly appealing for their inherent tunability both by chemical and physical modifications such as applied fields, alloying, heterostructuring and even moiré twist angle. In addition, controlling magnetism in such systems is a much easier feat than in bulk 3D materials where much more exchange pathways and interactions needed to be considered. Moreover, it is well-known that the combination of exchange fields and spin-orbit coupling in 2D hexagonal lattices gives rise to several topological phases~\cite{Niu2010, Yang2011, Qiao2011, Ezawa2011, Ren2016, Offidani2018, Hogl2018, Zou2020, Vila2021}. It is with these motivations that we here investigate how selective tuning of symmetry and magnetism can influence and control the resulting topology in a 2D magnetic system. Specifically, we select a monolayer of a known transition-metal phosphorus trichalcogenide (MnPSe$_3$)~\cite{Natalya2023}, and systematically perturb selected symmetries to explore the topological phases associated to different magnetic space groups and their corresponding symmetry protection. 

A special class of these are composite symmetries which are formed by combining two distinct symmetries, whether local or crystalline, in such a way that while each individual symmetry may be broken, the overall symmetry is preserved. Systems with such composite symmetries can therefore host topological phases while the individual symmetries constituting the composite symmetry are not preserved themselves, offering up new routes to exploring symmetry protected topological phases in matter. We identify such a composite symmetry (anti-unitary, order-two~\cite{shiozaki2014}) in MnPSe$_3$, $A=P\times(T\times M_y)$, where $P$, $T$, and $M_y$ respectively represent inversion, time-reversal, and mirror symmetry across the plane perpendicular to the $y$ axis. We find that combining this composite symmetry with three-fold rotation when the Mn ferromagnetic moments are in the out-of-plane direction stabilizes a semi-metallic phase. With this as our starting point, we investigate how by selectively perturbing the symmetries of the system, we can induce metal-insulator transitions and topological phase transitions. We explicitly demonstrate how these can be achieved in real systems through magnetization tilts and appropriate chemical substitutions. We do this by employing both an eight-band tight-binding Hamiltonian and a two-band effective $\mathbf{k}\cdot\mathbf{p}$ model, in addition to density functional theory (DFT) calculations.

The paper is organized as follows. In Sec.~\ref{sec:material}, we introduce the material system of interest, MnPSe$_3$. Then, in Sec.~\ref{sec:model}, we present our model and method. Subsequently, in Sec.~\ref{sec:orientation}, we analyze the electronic properties of MnPSe$_3$ for different magnetic orientations. In Sec.~\ref{sec:symmetry interplay}, we explore several symmetry-breaking designs and investigate their topological classification. We identify the band topology intrinsic to the band manifolds, which provides further insights into hosting a nontrivial bulk gap by introducing Rashba Spin-Orbit Coupling (SOC) and studying the Janus compound Mn$_2$P$_2$S$_3$Se$_3$. Finally, we summarize and conclude our findings in Sec~\ref{sec:conc}.

\section{\label{sec:material}Material system: \NoCaseChange{\MnPSe}}
\begin{figure}[]
    \centering
    \includegraphics[width=\columnwidth]{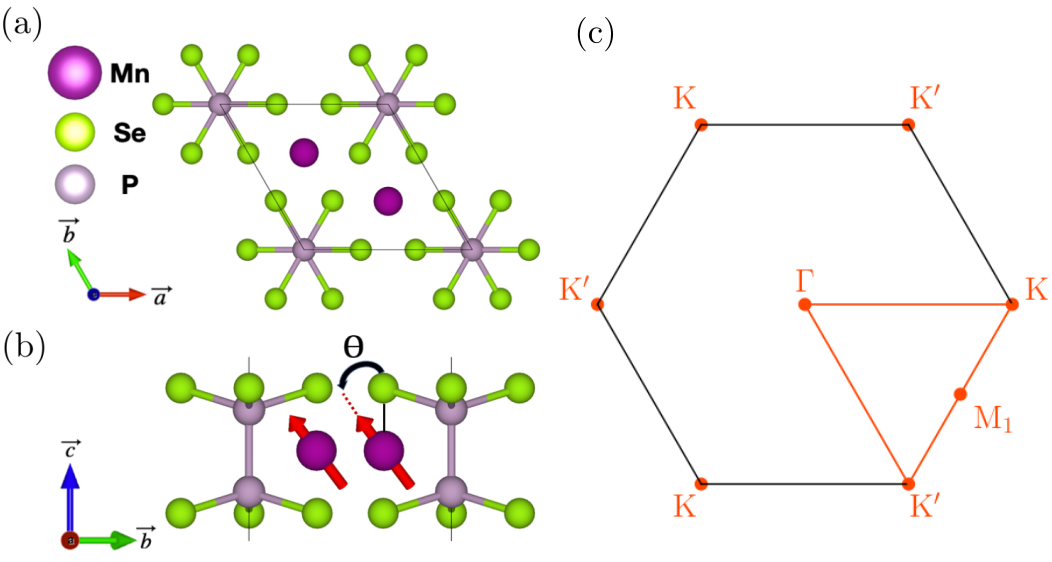}
    \caption{(a) Top view of a \MnPSe monolayer primitive cell. (b) We are studying the collinear FM magnetic order where the magnetic moment on the transition-metal Mn is oriented with the polar angle $\Theta$ to the layer-normal direction. (c) 2D Brillouin Zone (BZ) and the associated high-symmetry k-path with the associated high-symmetry k-points.}
    \label{fig:MnPSe3 Lattice Structures}
\end{figure}

Transition metal phosphorus trichalcogenides (TMPTs) are a family of hexagonal van der Waals materials with the general formula $M$P$X_3$ ($M$ = Mn, Fe, Co, Ni and $X$ = S, Se). Bulk TMPTs typically have a monoclinic structure ($C_2/m$, no. 12), except for MnPSe$_3$ and FePSe$_3$, which possess a rhombohedral structure ($R\bar{3}$, no. 148)~\cite{Wang2018}, composed of van der Waals bonded stacked monolayers. Our focus is on the isolated monolayer of MnPSe$_3$, presented in Fig.~\ref{fig:MnPSe3 Lattice Structures}. In TMPT monolayers, magnetic order usually exhibits in-plane antiferromagnetic (AFM) coupling, manifesting as either Néel or striped AFM order~\cite{Sivadas2015,Mak2019}. Due to the energetic stability of AFM ordering in TMPT monolayers under strain and pressure~\cite{Zhang2016,Macdonald2016}, less attention has been given to their ferromagnetic (FM) metastable state. In principle, FM ordering can be achieved by applying an external Zeeman field or, in some cases, through carrier doping~\cite{Li2014}. Electronic properties in such an FM-ordered state can undergo significant changes, resulting in new topological features that are not present in the AFM-ordered ground state~\cite{Natalya2023}. We note that a comprehensive study of the evolution of the magnetic exchange parameters was carried out previously~\cite{Pei2018}, reproducing well the reported AFM ground state and electronic structure.
%
\section{\label{sec:model}Model and method}
First-principles calculations based on DFT were carried out using the Vienna \textit{ab initio} simulation package (VASP)~\cite{KohnSham1965,Kresse1996,Kresse1999} for electronic band structure calculations. The Perdew-Burke-Ernzerhof (PBE) generalized gradient approximation (GGA) was adopted for the exchange-correlation potential~\cite{Perdew1997} and the wave functions and pseudopotentials were generated within the projector-augmented wave (PAW) method~\cite{Blochl1994}. 
The 2D monolayer was simulated by considering a sufficiently thick vacuum layer and the collinear FM order of the magnetic moment on Mn with SOC was taken into account (see Appendix.~\ref{app:DFT details}). We find minimal structural differences in the fully relaxed lattice parameters for different magnetic orientations, therefore we fix the structural parameters for all magnetic orientations considered. The topology of the calculated band structures was analyzed via the explicit calculation of the Berry curvature and the Chern number. For the latter, maximally localized Wannier functions (MLWFs) were constructed using the Wannier90 code~\cite{Souza2001,Wang2007,Marzari2012}. 

To study the electronic bands near the Fermi level, we construct a tight-binding Hamiltonian for the monolayer honeycomb lattice, comprising two Mn d orbitals: $d_{xz}$ and $d_{yz}$. The resulting Hamiltonian can be expressed as
\begin{align} \label{eq:TB_Hamiltonian_MnPSe3}
    \hat{H}=&{\sum_{\left \langle i,j \right \rangle,a,b,\alpha}(t_{i,j})_{a,b}\hat{c}^{\dag}_{i,a,\alpha}\hat{c}_{j,b,\beta}} \\ 
            &+{\lambda_{SO}\sum_{i,(a,\alpha);(b,\beta)}{\tau_y}_{a,b}{s_z}_{\alpha,\beta}\hat{c}^{\dag}_{i,a,\alpha}\hat{c}_{i,b,\beta}} \nonumber \\
            &+{i\lambda_{R}\sum_{\left \langle i,j \right \rangle,a,\alpha,\beta}\left [(\mathit{\mathbf{s}}_{\alpha,\beta}\times\hat{\mathbf{d}}_{i,j})\cdot\hat{z}\right]\hat{c}^{\dag}_{i,a,\alpha}\hat{c}_{i,a,\beta}} \nonumber \\
            &+{M\sum_{i,a,\alpha,\beta}(\mathit{\mathbf{\hat{m}}}\cdot \mathit{\mathbf{S}}_{\alpha,\beta})\hat{c}^{\dag}_{i,a,\alpha}\hat{c}_{i,a,\beta}}, \nonumber 
\end{align}
where $\hat{c}^{(\dag)}_{i}$ is the electron annihilation (creation) operator on site $i$ with Pauli matrices $\tau$ and $s$ acting on the orbital and spin, respectively. The angular bracket in $\left \langle i,j \right \rangle$ represents nearest-neighbor (NN) sites, and $\hat{\mathbf{d}}_{i,j}$ is the unit vector connecting site $j$ to site $i$. The SOC strength is represented by $\lambda_{SO}$ for intrinsic SOC and $\lambda_{R}$ for extrinsic Rashba SOC, arising only when inversion symmetry is absent \cite{Kane2005, Gmitra2009, Konschuh2010}. The last term represents the uniform exchange field, with strength $M$ and orientation $\mathit{\mathbf{\hat{m}}}=(\sin\Theta,0,\cos\Theta)$ determined by the polar angle $\Theta$ in Fig.~\ref{fig:MnPSe3 Lattice Structures}(b). Details on how the NN hopping matrix $t_{i,j}$ is set and the symmetries in the Hamiltonian with their representations in the orbital and spin bases are provided in Appendix.~\ref{app:TB details}. 

\section{\label{sec:results}Results}

\subsection{\label{sec:orientation} Magnetic moment orientation dependent electronic properties}
\begin{figure}[]
    \centering
    \includegraphics[width=1.01\columnwidth]{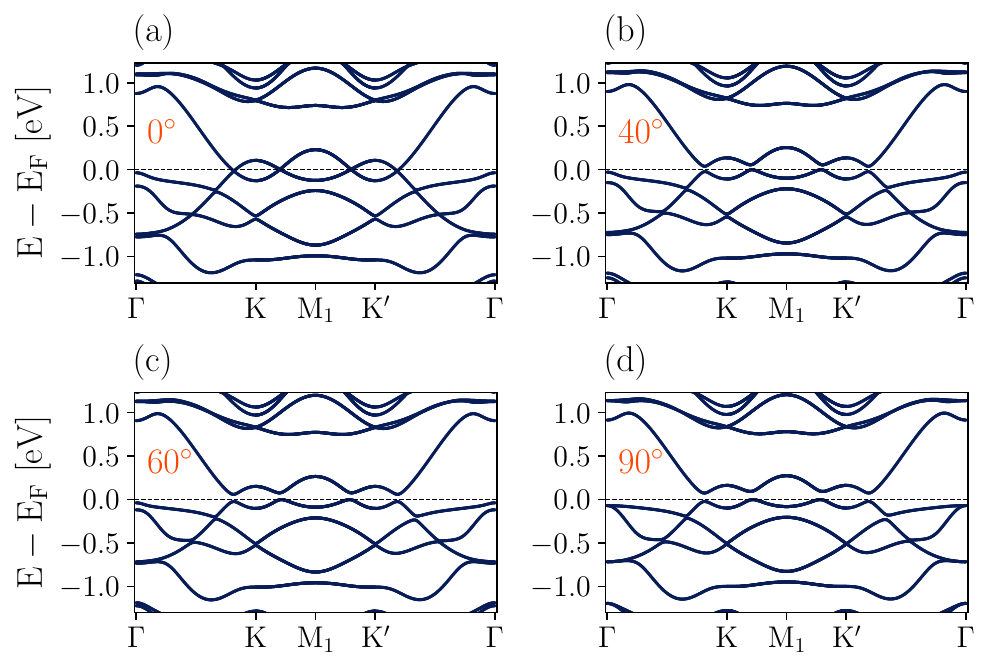}
    \caption{DFT calculations of the electronic structure of MnPSe$_3$ with SOC for different spin orientations. The Fermi level is marked by the dashed line in all plots and is set to 0 eV. (a) For $\Theta=0^{\circ}$ (FM$_z$), we find a band crossing along $\Gamma$ to $\rm K$ and a small gap along $\rm K$ to $\rm M_1$ near the Fermi level, as shown in more detail in Fig.~\ref{fig:Band and kp}(a). This topologically stable band crossing is opened up resulting in a trivial gap when $\Theta\neq0^{\circ}$ (see Sec.~\ref{sec:symmetry analysis of non-zero polar angle}). The gap size becomes larger as $\Theta$ increases. (b) $\Theta=40^{\circ}$, (c) $\Theta=60^{\circ}$, and (d) $\Theta=90^{\circ}$.}
    \label{fig:MnPSe3 Magnetic orientation band structures}
\end{figure} 
We explore the relativistic electronic band structure of the ferromagnetic (FM) \MnPSe~upon varying the orientation of its magnetic moment on Mn. This orientation is described by the polar angle $\Theta$, which range from the out-of-plane FM$_z$ configuration ($\Theta=0^\circ$) to the in-plane FM$_x$ configuration ($\Theta=90^\circ$). In the FM$_z$ case, we observe a band crossing from $\Gamma$ to $\rm K$ and a small gap from $\rm K$ to $\rm M_1$ near the Fermi level, as shown in Fig.~\ref{fig:MnPSe3 Magnetic orientation band structures}(a) and Fig.~\ref{fig:Band and kp}(a). When the magnetization is tilted in-plane ($\Theta\neq0^{\circ}$), this crossing is opened up resulting in a direct band gap. Fig.~\ref{fig:MnPSe3 direct gap, indirect gap} illustrates both the direct and indirect band gaps ($\Delta_\text{min}$) as a function of the tilting angle. We emphasize that for small values of $\Theta$, although the band crossing from $\Gamma$ to $\rm K$ becomes gapped, \MnPSe~exhibits a metallic behavior due to small electron pockets. However, as the Fermi level crosses the global indirect gap when $\Theta>\Theta_0\approx10^\circ$, \MnPSe~becomes an insulator. Furthermore, by increasing $\Theta$ up to $\Theta = 90^\circ$, the direct (indirect) band gap reaches a maximum value of $94$ ($75$) meV. 
\begin{figure}[]
    \centering
    \includegraphics[width=0.88\columnwidth]{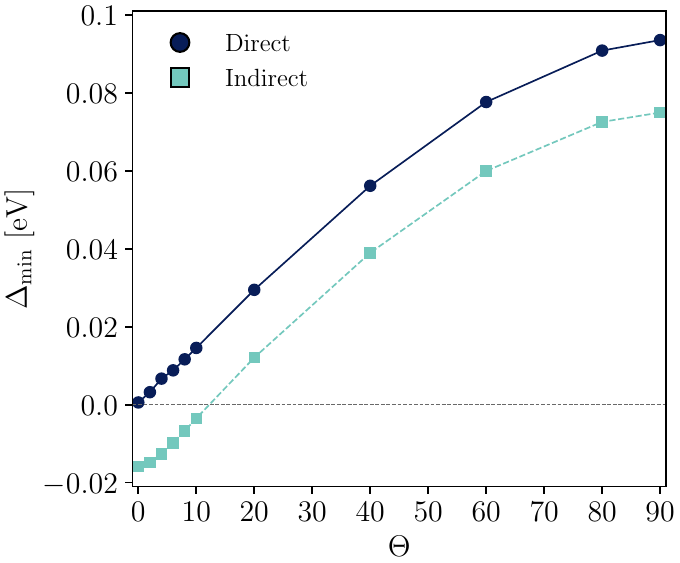}
    \caption{Calculated direct and indirect band gaps values as a function of the Mn magnetization orientation, where $\Theta=0^{\circ}$ corresponds to Mn spin in the out-of-plane direction, and $\Theta=90^{\circ}$ is in-plane. The value of the indirect band gap is negative when the top of the valence band is higher than the bottom of the conduction band in energy. \MnPSe~is a metal with small electron pockets for FM$_z$ and small values of polar angles $\Theta$, which becomes an insulator when the Fermi level is passing through the band gap for $\Theta>\Theta_0\cong10^{\circ}$.}
    \label{fig:MnPSe3 direct gap, indirect gap}
\end{figure} 
To explain the band crossing from $\Gamma$ to $\rm K$ and the gap opening upon tilting the magnetization, we perform a symmetry analysis of \MnPSe~in the next subsections. The results of this analysis are also summarized in Table~\ref{tab:Symmetry Analysis table}(a) and (b). Subsequently, we proceed to construct a $\mathbf{k}\cdot\mathbf{p}$ Hamiltonian to elucidate the stability of the protected crossing by symmetries.

\subsubsection{\label{sec:symmetry analysis of FM_z}Symmetry of \MnPSe~with FM$_z$}
\begin{figure}[]
    \centering
    \includegraphics[width=1.01\columnwidth]{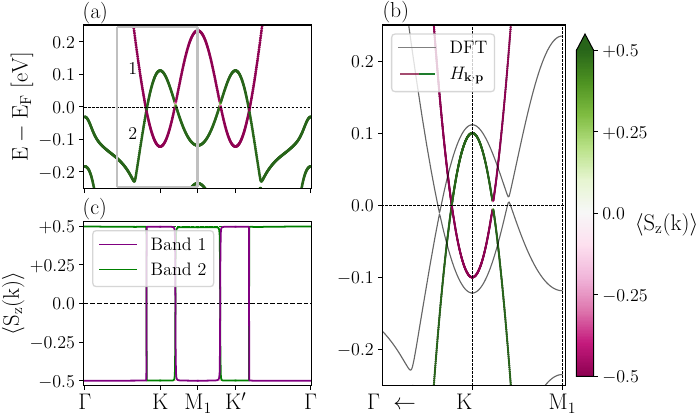}
    \caption{(a) Low-energy DFT band structure for $\Theta=0^{\circ}$ (FM$_z$), showing a band crossing along $\Gamma$ to $\rm K$ that is protected by the order-two antiunitary composite symmetry $A$ and the rotation symmetry $C_{3z}$. However, along $\rm K$ to $\rm M_1$ ($k_y\neq0$), the Hamiltonian does not commute with $A$, leading to a small gap. (b) The corresponding $H_{\mathbf{k}\cdot\mathbf{p}}$ captures the relevant features of the DFT bands with the band crossing along $\Gamma$ to $\rm K$ and the gap opening along $\rm K$ to $\rm M_1$. The region plotted in (b) corresponds to the grey box indicated in (a). The green and pink coloring of the  $H_{\mathbf{k}\cdot\mathbf{p}}$ represent the spin projected along $z$.  (c) The calculated spin texture (in units of $\hbar$) of two bands near the Fermi level along the high symmetry k-path suggests that a small spin-mixing contributes to the gap opening along $\rm K$ to $\rm M_1$, but there is no spin-mixing at the band crossing $\mathbf{k}_0^{*}$ along $\Gamma$ to $\rm K$. In all plots the Fermi level is set to 0 eV and is marked by a dashed line.}
    \label{fig:Band and kp}
\end{figure}

The magnetic space group of FM$_z$ corresponds to P-3(147.13), which includes three-fold rotation symmetry $C_{3z}$ and inversion symmetry $P$~\cite{Aroyo2011}, as shown in Table.~\ref{tab:Symmetry Analysis table}(a). While time reversal symmetry $T$ is explicitly broken due to the intrinsic magnetic moments, an order-two composite antiunitary symmetry $A$ can be defined as $A=P\times(T\times M_y)$, with $A^2=\mathbbm{1}$, where $M_y$ is the mirror symmetry across the mirror plane perpendicular to the $y$ axis. In momentum representation, $P$, $C_{3z}$ and $A$ symmetries are represented as 
\begin{align} \label{eq:momentum representation of symmetries}
    &U_PH(\mathbf{k})U_P^{\dag}=H(-\mathbf{k}), U_P^2=\mathbbm{1}, \nonumber \\
    &U_{3z}H(\mathbf{k})U_{3z}^{\dag}=H(R_{3z}\mathbf{k}), U_{3z}^3=-\mathbbm{1}, \nonumber \\
    &AH(\mathbf{k})A^{-1}=H(k_x,-k_y), A=U_{A}K, U_{A}U_{A}^{*}=\mathbbm{1} 
\end{align}
where $R_{3z}$ is a three-fold rotation matrix in 2D momentum space and $K$ denotes complex conjugation. Along $k_y=0$, the Hamiltonian commutes with $A$, which imposes the band crossing at $\mathbf{k}_0^{*}$ along $\Gamma$ to $\rm K$ ($\mathbf{k}_y=0$) in Fig.~\ref{fig:Band and kp}(a), in addition to two other band crossings at $\mathbf{k}_1^{*}=\mathbf{K}+R_{3z}(\mathbf{k}_0^{*}-\mathbf{K})$ and $\mathbf{k}_2^{*}=\mathbf{K}+R_{3z}^{-1}(\mathbf{k}_0^{*}-\mathbf{K})$. 
We attribute the band crossing at $\mathbf{k}_0^{*}$ to the presence of the antiunitary symmetry $A$ ($A^2=\mathbbm{1}$) and the discrete rotation symmetry $C_{3z}$ ($ C_{3z}^3=-\mathbbm{1}$). The combination of these symmetries gives rise to a symmetry-protected accidental SM phase that exhibits topological stability~\cite{Chiu2016,Andreas2020}. It is important to note that this differs from a symmetry-enforced SM phase, which is protected by nonsymmorphic space groups~\cite{Young2015}. In our case, the given symmetries (symmorphic) do not mandate the existence of the band crossing, but rather explain the stability of the crossing once it occurs. 

\subsubsection{\label{sec:symmetry analysis of non-zero polar angle}Symmetry of \MnPSe~with FM$_{\Theta\neq0^{\circ}}$}
When the magnetization of the crystal is not aligned with the $z$-axis (the polar angle $\Theta$ of the magnetization is non-zero), the magnetic space group is reduced because the rotation symmetry $C_{3z}$ is broken, while the symmetry $P$ and $A$ are still preserved, as indicated in Table~\ref{tab:Symmetry Analysis table}(b). The breaking of the $C_{3z}$ symmetry leads to the opening of the band gap at $\mathbf{k}_0^{*}$. This is further supported by the following two-band effective $\mathbf{k}\cdot\mathbf{p}$ Hamiltonian, which forbids the band crossing in the absence of $C_{3z}$ symmetry.    

\subsubsection{\label{sec:kp Hamiltonian}$\mathbf{k}\cdot\mathbf{p}$ Hamiltonian construction}
We construct the two-band $\mathbf{k}\cdot\mathbf{p}$ Hamiltonian $H_{\mathbf{k}\cdot\mathbf{p}}(\tilde{\mathbf{k}})$ near the high-symmetry point $\rm K$, where $\tilde{\mathbf{k}}=\mathbf{k}-\mathbf{K}$, to explain three band crossings near $\rm K$ that are protected by both $A$ and $C_{3z}$ symmetries. We generate symmetry-preserved bases in matrix representations of rank-$2$ (spin degree of freedom) using the Qsymm package~\cite{Varjas2018} and express the Hamiltonian as 
\begin{align} \label{eq:H_kp construction}
    H_{\mathbf{k}\cdot\mathbf{p}}=&\alpha\cdot\left [k_x\sigma_x+k_y\sigma_y\right ]+\beta\cdot\left[\frac{(k_x^2-k_y^2)}{2}\sigma_x-k_xk_y\sigma_y\right ] \nonumber \\ 
    &+m_{s}\cdot\left[\frac{(k_x^2+k_y^2)}{2}\sigma_z\right]-B_z\cdot\sigma_z, 
\end{align}
where the redefined momenta centered at $\rm K$ are denoted as $k_x$ and $k_y$, after the renaming $\tilde{k}_i \rightarrow k_i$. The eigenvalues $\lambda_{1,2}$ of $H_{\mathbf{k}\cdot\mathbf{p}}$ in Eq.~\eqref{eq:H_kp construction} at $k_y=0$, are given by
\begin{equation} \label{eq:H_kp solution}
    \lambda_{1,2}^2=\left (\alpha k_x+\beta k_x^2/2 \right )^2+\left (-B_z+m_sk_x^2\right )^2,    
\end{equation}
where the first and second parenthesis term correspond to the contributions from off-diagonal spin mixing and spin splitting, respectively. The $\lambda_{1,2}=0$ is satisfied when $(-2\alpha/\beta)^2=B_z/m_s$, determining the location of the band crossing at $\mathbf{k}^{*}_0=(-2\alpha/\beta,0)$. This point lies along $\Gamma$ to $\rm K$ and is displayed in Fig.~\ref{fig:Band and kp}(b) for the parameter values $m_s=-1.3$, $B_z=-0.1$, and $\alpha=0.01$. 

From $\rm K$ to $\rm M_1$ ($k_y\neq0$), the symmetry $A$ no longer commutes with the Hamiltonian in Eq.~\eqref{eq:momentum representation of symmetries}. This leads to the generation of additional symmetry-allowed bases and extra spin mixing terms $(\alpha^{\prime} k_x)^2\sigma_x$ and $(\alpha^{\prime\prime} k_y)^2\sigma_x$ in Eq.~\eqref{eq:H_kp construction}. The spin texture along $\rm K$ to $\rm M_1$ in Fig.~\ref{fig:Band and kp}(c) illustrates the contribution of spin-mixing to the gap opening, as seen by the gradual change in spin texture between bands, as opposed to the much more abrupt change observed along $\Gamma$ to $\rm K$. Consequently, the gap forms along $\rm K$ to $\rm M_1$, as shown in Fig.~\ref{fig:Band and kp}(b). 

In the absence of $A$, the eigenvalues $\lambda_{1,2}$ of $H_{\mathbf{k}\cdot\mathbf{p}}(\tilde{\mathbf{k}})$ at $k_y=0$ are determined by $\lambda_{1,2}^2=\left (\alpha k_x+\beta k_x^2/2 \right )^2+(\alpha^{\prime} k_x)^2+\left (-B_z+m_sk_x^2\right )^2$. These eigenvalues cannot satisfy $\lambda_{1,2}^2=0$ for any non-zero value of $\alpha^{\prime}$. Therefore, when the symmetry $A$ is lifted, the band crossing is not protected, resulting in the opening of a gap along $\Gamma$ to $\rm K$ and explaining why the gap crossing does not occur along the $\rm K$ to $\rm M_1$ path. 

\begin{table*}[]
\caption{\label{tab:Symmetry Analysis table}Symmetry designs and identification of each topological phase}
 \centering
 \begin{ruledtabular}
 \begin{tabular}{|| l | c | c | c | c | l ||}
 {\textbf{Symmetry~Design}} & {$P$} & {$C_{3z}$} & {$A=P\times(T\times M_y)$} & {\textbf{Classification}} & {\textbf{Topological~Phases}}\\  
 \hline\hline                    
 (a) MnPSe$_3$~($\rm FM_z$) & $\textcolor{teal}{\checkmark}$ & $\textcolor{teal}{\checkmark}$  & $\textcolor{teal}{\checkmark}$ & N/A\footnote{The classification presented in the table corresponds to the classification of gapped phases. If we consider gapped phases, both the first Chern character $\mathbb{Z}$ ($C_1$) and the independent representation rings over $\mathbb{Z}$ completely specify the stable first-order topological phases.} & Symmetry-Protected Accidental SM\\ 
 \hline
 (b) MnPSe$_3$~($\rm FM_{\Theta\neq0^{\circ}}$) & $\textcolor{teal}{\checkmark}$ & $\times$ & $\textcolor{teal}{\checkmark}$ & $\mathbb{Z}\oplus\mathbb{Z}$ & Trivial~Gap~($C_1=0$) \\
 \hline
 (c) MnPSe$_3$~(Modified Se positions,~$\rm FM_z$) & $\times$ &  $\textcolor{teal}{\checkmark}$ & $\times$ & $\mathbb{Z}$ & Nontrivial~Gap~($C_1 = -1$) \\
 \hline
 (d) MnPS$_{1.5}$Se$_{1.5}$~(Rashba~SOC,~$\rm FM_z$) & $\times$ &  $\textcolor{teal}{\checkmark}$ & $\times$ & $\mathbb{Z}$ & Nontrivial~Gap~($C_1 = 2$) \\
 \hline
 (e) MnPS$_{1.5}$Se$_{1.5}$~(Rashba~SOC,~$\rm FM_{\Theta<\Theta_c}$) & $\times$ &  $\times$ & $\times$ & $\mathbb{Z}$ & Nontrivial~Gap~($C_1 = 2$) \\
 \hline
 (f) MnPS$_{1.5}$Se$_{1.5}$~(Rashba~SOC,~$\rm FM_{\Theta>\Theta_c}$) & $\times$ &  $\times$ & $\times$ & $\mathbb{Z}$ & Trivial~Gap~($C_1 = 0$) \\
\end{tabular}
\end{ruledtabular}
\end{table*}

Similarly, without $C_{3z}$, the form of the Hamiltonian $H_{\mathbf{k}\cdot\mathbf{p}}$ in Eq.~\eqref{eq:H_kp construction} changes, as more symmetry-allowed bases are generated. At $k_y=0$, the eigenvalues of the Hamiltonian are given by $\lambda_{1,2}^2=(\alpha k_x+\beta k_x^2/2)^2+(-B_z+m_s^{\prime}k_x+m_s k_x^2)^2$, which is different from the form in Eq.~\eqref{eq:H_kp solution} due to the presence of the added spin splitting term ($m_s^{\prime}k_x$) linear in $\mathbf{k}$. This change in the Hamiltonian also forbids the band crossing at $\mathbf{k}_0^{*}$ along the path from $\Gamma$ to $\rm K$ at FM$_{\Theta\neq0^{\circ}}$, thus explaining the gap opening when the magnetization is tilted. Therefore, it is clear that both symmetries $A$ and $C_{3z}$ are required to protect the band crossing, otherwise a gap can be opened, giving us a wealth of options to explain how topological phases can be controlled with symmetry.

\subsection{\label{sec:symmetry interplay}Topology of symmetry designs}
We design several settings with distinct broken symmetries of \MnPSe~and its isostrutural Janus compound Mn$_2$P$_2$S$_3$Se$_3$ in order to identify and classify topological phases. The details of these settings are summarized in Table~\ref{tab:Symmetry Analysis table}.        
\subsubsection{\label{sec:topological classification}Topological classification}
As we discussed in the Sec.~\ref{sec:symmetry analysis of FM_z}, in the presence of both symmetries $C_{3z}$ and $A$ for FM$_z$, the crossings of two bands near the Fermi level at $\mathbf{k}_0^{*}$ along $\Gamma$ to $\rm K$ with two other points $\mathbf{k}_{1,2}^{*}$ connected through the $R_{3z}$ rotation matrix, are symmetry-protected. However, in the Sec.~\ref{sec:symmetry analysis of non-zero polar angle}, when the polar angle of the magnetization is non-zero (broken $C_{3z}$), the crossing is no longer protected, and a finite gap is opened, as shown in Fig.~\ref{fig:MnPSe3 Magnetic orientation band structures}. We now analyze the topological features of this gap. Despite the Fermi level passing through the bands above and below the gap in a small energy window (see Fig.~\ref{fig:Band and kp}), making the system a SM (see Fig. \ref{fig:MnPSe3 direct gap, indirect gap} for small angles), we consider the band topology to be an intrinsic property of the band, independent of where the Fermi level sits in a particular system. We focus on the band manifolds below the gap, which can be adiabatically tuned to make the Fermi level cross through the gap without changing the gap’s topology as long as the gap remains open under symmetry-preserving perturbations. 

The $C_{3z}$ symmetry can protect higher-order topological phases that are classified as $\mathbb{Z}_3$~\cite{Benalcazar2018}, however in this study, our focus is on the strong index of the first-order topological phase and its classification. The first Chern character of the two-form of the band manifolds below the gap becomes a topological invariant known as the first Chern number $C_1$~\cite{Thouless1982,Dixiao2010,Chiu2016}. Additionally, a comprehensive classification of topological phases requires a set of independent integers (the representation rings over $\mathbb{Z}$ of the stabilizer groups at high-symmetry $k$ points) that specify the representation of band manifolds satisfying the compatibility relations~\cite{Kruthoff2017,Po2020}. This classification can be obtained computing the $G$-equivariant $K$-theory constrained by space group symmetry $G$~\cite{Freed2013}. The classification $\mathbb{Z}$ presented in Table.~\ref{tab:Symmetry Analysis table}(c) and (d) corresponds to the first Chern character, and we do not focus on an in-depth study of representation rings in our study.

In the presence of both order-two spatial symmetries $A$ ($A^2=\mathbbm{1}$) and $P$ ($P^2=\mathbbm{1}$) satisfying $\comm{A}{P}=0$ in our system, the topological classification corresponds to $K^{U}_{\mathbb{R}}(s=0,t=0;\delta=0,\delta_{\parallel}=0)=\pi_0(R_0)\oplus\pi_0(R_0)=\mathbb{Z}\oplus\mathbb{Z}$ from the K-theory~\cite{shiozaki2014}, as presented in Table.~\ref{tab:Symmetry Analysis table}(b). In this case, the first invariant $\mathbb{Z}$ represents the first Chern number $C_1$, while the other invariant $\mathbb{Z}$ is given by $[\Gamma_{0,\infty}]=N(\Gamma^{+}_0)-N(\Gamma^{+}_{\infty})=-(N(\Gamma^{-}_0)-N(\Gamma^{-}_{\infty}))$. Here, $N(\Gamma^{\pm}_{i})$ denotes the number of states below the gap in the two subsectors with different parity $\pm$ of $H_{\pm}(\Gamma_i)$ at the inversion-invariant $k$ points $\Gamma_0=(0,0)$ and $\Gamma_{\infty}=\infty$. The relation $(-1)^{C_1}=(-1)^{[\Gamma_{0,\infty}]}$~\cite{Hughes2011,Turner2012} holds between $C_1$ and $[\Gamma_{0,\infty}]$, and the Kane-Mele $Z_2$ index $\nu$~\cite{Fu2007} is given by $\nu=C_1\mod2=[\Gamma_{0,\infty}]\mod2$.

\begin{figure*}[]
    \centering
    \includegraphics[width=\textwidth]{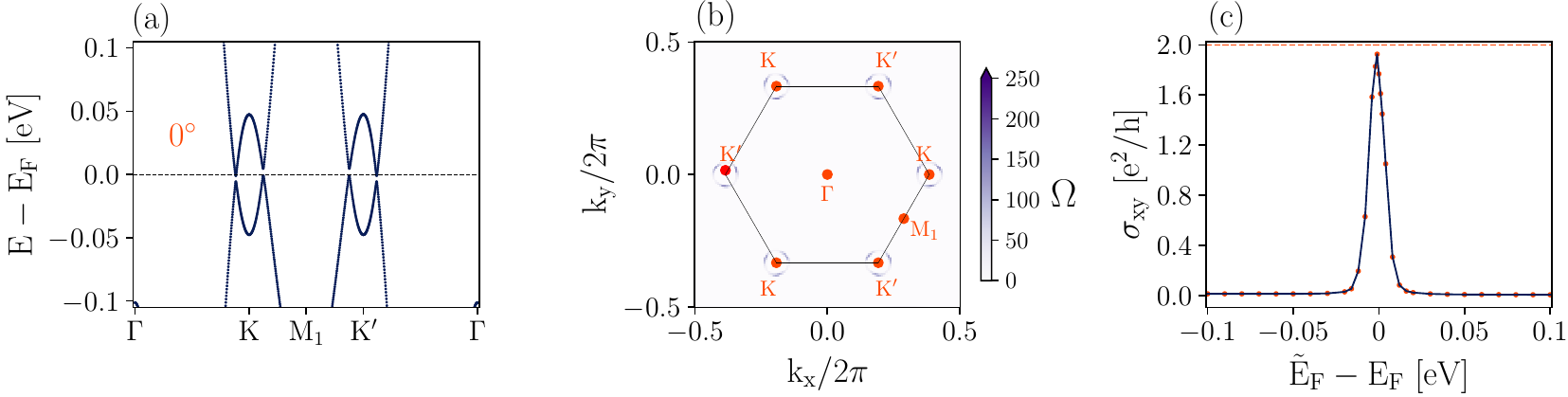}
    \caption{ DFT calculations of the electronic and topological properties of the Janus compound  Mn$_2$P$_2$S$_3$Se$_3$ with the Mn moment out-of-plane $\Theta\ = 0^{\circ}$. (a) The calculated electronic structure shows that the band along  $\Gamma$ to $\rm K$, which was previously symmetry-protected in MnPSe$_3$, is now open in Mn$_2$P$_2$S$_3$Se$_3$ (see Table.~\ref{tab:Symmetry Analysis table}). This is due to the breaking of the inversion symmetry $P$, which results in the breaking of the symmetry $A$. (b) Calculated  $\Omega(\mathbf{k})$, the trace of the Berry curvature, in units of $e^2/h$. Integrating $\Omega$ over the first BZ, gives the Chern number, $C_1=2$ (see Eq.~\eqref{eq:Chern number calculation}). (c) Calculated anomalous Hall conductivity $\sigma_{xy}(\rm{\tilde{E}_F})$ for various values of the Fermi level $\rm \tilde{E}_F$ in the low temperature limit (T$=0$). For FM$_z$ at $\rm \tilde{E}_F=E_F$, the system is a metal due to the presence of small electron pockets in the band above the gap. This leads to the value of $\sigma_{xy}(\rm{\tilde{E}_F})$ being not as quantized as $C_1=2$ without the Hall plateau.}
    \label{fig:Mn2P2S3Se3 FM_z}
\end{figure*}

\subsubsection{\label{sec:topological phases}Identification of topological phases}
We construct the Berry connection $A(\mathbf{k})$ of the band manifolds $\left \{ \ket{\Psi^{a}(\mathbf{k})} \right \}^{N}_{a=1}$ below the energy gap as a 1-form with values in Lie algebra $U(N)$, expressed as $A(\mathbf{k})^{ab}_{i}=i\bra{\Psi^{a}(\mathbf{k})}\ket{\partial_{k_i}\Psi^{b}(\mathbf{k})}$, where $a$ and $b$ are band indices ranging from $1$ to $N$. The Berry curvature $F(\mathbf{k})$ is defined as a 2-form $F=dA+A\wedge A$ with values in $U(N)$, given by $F^{ab}_{ij}$. We can define a $U(1)$ gauge-invariant object $\Omega$ as 
\begin{align} \label{eq:U(1) Berry curvature}
    &\Omega=\Omega_{ij}dk^i\wedge dk^j=\Tr(F), \nonumber \\
    &\Omega(\mathbf{k})_{i,j}=F(\mathbf{k})^{aa}_{i,j}, \nonumber \\
    &F(\mathbf{k})^{aa}_{i,j}=\bra{\partial_{k_i}\Psi^{a}(\mathbf{k})}\ket{i\partial_{k_j}\Psi^{a}(\mathbf{k})}-\bra{\partial_{k_j}\Psi^{a}(\mathbf{k})}\ket{i\partial_{k_i}\Psi^{a}(\mathbf{k})}. 
\end{align}
To calculate the first Chern number, we evaluate the integral of $\Omega$ over the first BZ as
\begin{equation} \label{eq:Chern number calculation}
    C_1=\frac{1}{2\pi}\int_{BZ}\Omega(\mathbf{k}),
\end{equation} 
where it takes the form $C_1=1/(2\pi)\sum_{\mathbf{k}}\Omega(\mathbf{k})$ in the finite $k$-grid expression~\cite{Fukui2005,Vanderbilt2006}. In the case of \MnPSe~with a non-zero polar angle $\Theta$, the $C_{3z}$ symmetry is broken, resulting in the opening of a finite gap with $C_1=0$ as shown in Table~\ref{tab:Symmetry Analysis table}(b).

\begin{figure}[b]
    \centering
    \includegraphics[width=0.95\columnwidth]{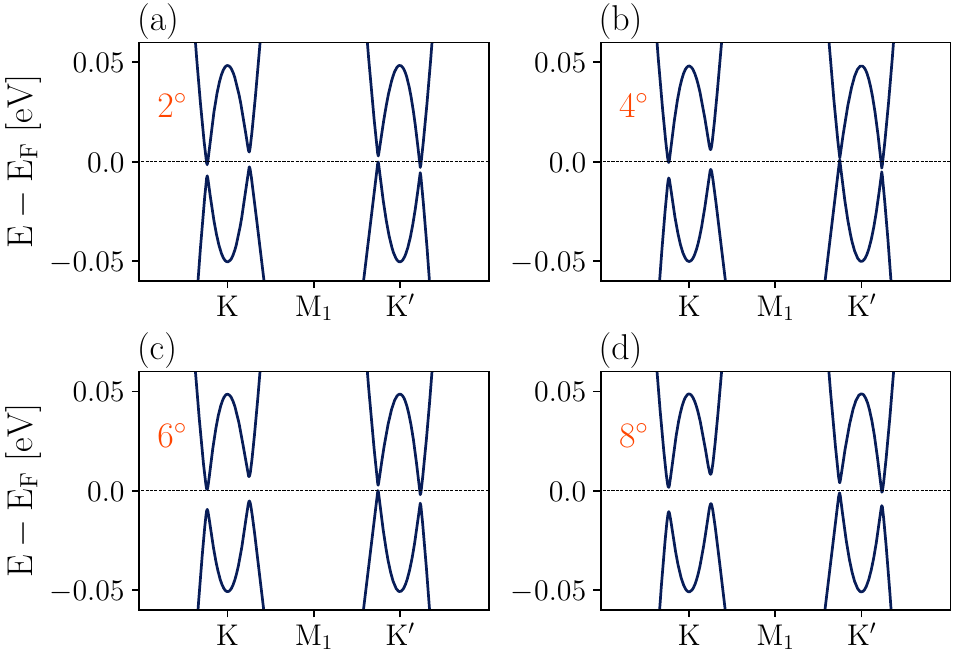}
    \caption{DFT calculations of the electronic structure of Mn$_2$P$_2$S$_3$Se$_3$ with SOC for different Mn spin orientations. The Fermi level is marked by the dashed line in all plots and is set to 0 eV.  A topological phase transition occurs with gap closure at the critical polar angle $\Theta\cong5.5^{\circ}$ along the $\rm M_1$ to $\rm K^{\prime}$, changing the first Chern number from $C_1=2$ to $C_1=0$. Figures (a)-(d) show the band structure for different polar angles: (a) $\Theta=2^{\circ}$, (b) $\Theta=4^{\circ}$ $(C_1=2)$, (c) $\Theta=6^{\circ}$, and (d) $\Theta=8^{\circ}$ $(C_1=0)$.}
    \label{fig:Mn2P2S3Se3 orientation dependent band}
\end{figure} 

\subsubsection{\label{sec:MnPSe3 broken inversion}\MnPSe~with crystal structure distortions: Table.~\ref{tab:Symmetry Analysis table}(c)}
We can consider a method to intentionally break specific symmetries by distorting the lattice, thereby disrupting the symmetries. For example, rotating all three top Se atoms by $1^{\circ}$ around the axis passing through Mn manually breaks the inversion symmetry (and consequently $A$), while preserving the $C_{3z}$ symmetry. In this particular configuration, a topologically nontrivial gap opens with $C_1=-1$, as presented in Table~\ref{tab:Symmetry Analysis table}(c).

\begin{figure*}[]
    \centering
    \includegraphics[width=0.98\textwidth]{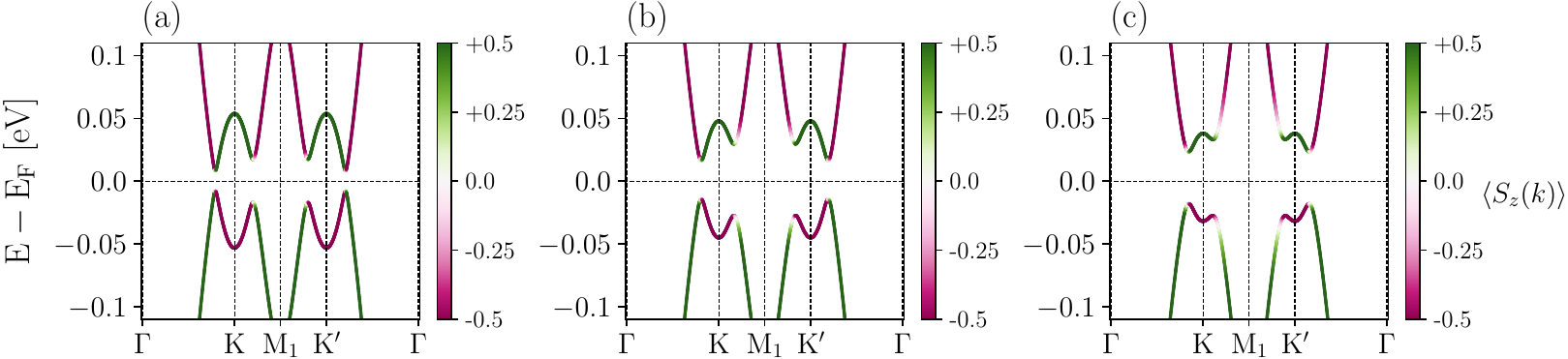}
    \caption{Tight-binding band structures and spin textures (in units of $\hbar$) of Mn$_2$P$_2$S$_3$Se$_3$ showing the effect of increasing the Rashba SOC strength, $\lambda_R$, on the size of the band gap. The band structures for different values of $\lambda_R$ are calculated for FM$_z (\Theta=0^{\circ})$ using the tight-binding Hamiltonian in Eq.~\eqref{eq:TB_Hamiltonian_MnPSe3} with parameters $\lambda_{SO}/t_{x,x}=0.2$, $M/t_{x,x}=-2.6$, $t_{y,y}/t_{x,x}=0.1$, and  $t_{x,y}=t_{y,x}=\sqrt{3}/2\cdot(t_{x,x}-t_{y,y})$ where $t_{x,x}=0.52$. Figures (a)-(c) correspond to values of $\lambda_R$: (a)$\lambda_R/\lambda_{SO}=0.25$, (b) $\lambda_R/\lambda_{SO}=0.5$, and (c) $\lambda_R/\lambda_{SO}=0.75$. }
    \label{fig:Mn2P2S3Se3 Rashba SOC strength}
\end{figure*} 

\subsubsection{\label{sec:MnPS3Se3 before the transition}Janus Mn$_2$P$_2$S$_3$Se$_3$ with FM$_{\Theta<\Theta_c}$: Table~\ref{tab:Symmetry Analysis table}(d) and (e)}
To implement a more realistic way of breaking inversion symmetry, we calculate the band structure of the Janus compound Mn$_2$P$_2$S$_3$Se$_3$, which naturally breaks inversion symmetry due to the replacement of the top Se atoms by S. The resulting bands at FM$_z$ are shown in Fig.~\ref{fig:Mn2P2S3Se3 FM_z}(a). Again, the Fermi level accidentally passes through bands due to the small band gap, but the gap can still be classified by the $C_1$ invariant of the band manifolds below the gap. The gauge-invariant $\Omega(\mathbf{k})$ of the manifolds is displayed in Fig.~\ref{fig:Mn2P2S3Se3 FM_z}(b), yielding $C_1=2$, confirming the nontrivial band topology of this material. 

We calculate the anomalous transverse Hall conductivity $\sigma_{xy}$ by integrating the gauge-invariant $\Omega(\mathbf{k})$ over the first BZ up to the given Fermi level $\rm \tilde{E}_F$ with the occupation given by the Fermi-Dirac factor $f_{FD}(\tilde{\textrm{E}}_\textrm{F},\mathbf{k})$ as $\sigma_{xy}(\tilde{\textrm{E}}_\textrm{F})=e^2/h\int_{BZ}f_{FD}(\tilde{\textrm{E}}_\textrm{F},\mathbf{k})\Omega(\mathbf{k})$. In the low temperature limit (T$=0$), this expression can be written using a finite $k$-grid as 
\begin{equation} \label{eq:anomalous transverse Hall(E)}
    \sigma_{xy}(\tilde{\textrm{E}}_\textrm{F})=\frac{e^2}{h}\sum_{\mathbf{k}}\Tr_{N_{\mathbf{k}}}[F(\mathbf{k})],
\end{equation}
where the trace of the 2-form Berry curvature $F(\mathbf{k})$ is taken over the $N_{\mathbf{k}}$ occupied states at each $\mathbf{k}$-point up to the given Fermi level $\rm \tilde{E}_F$. The calculated $\sigma_{xy}(\rm \tilde{E}_F)$ is shown in Fig.~\ref{fig:Mn2P2S3Se3 FM_z}(c), where the value of the Hall conductivity reaches almost $\sigma_{xy}=2\cdot e^2/h$ near the gap region due to the system being metallic, as there exist very small electron pockets. With a non-zero polar angle $\Theta$ that breaks $C_{3z}$ symmetry, its phase remains topologically nontrivial until the gap closes at $\Theta = \Theta_c$.

\subsubsection{\label{sec:MnPS3Se3 after the transition}Janus Mn$_2$P$_2$S$_3$Se$_3$ with FM$_{\Theta>\Theta_c}$: Table~\ref{tab:Symmetry Analysis table}(f)}
Moreover, we observe a topological phase transition with the gap closing along $\rm M_1$ to $\rm K^{\prime}$ at the critical angle $\Theta_c\cong5.5^{\circ}$ from $C_1=2$ to $C_1=0$ by tuning the polar angle $\Theta$ of the magnetic orientation (see Fig.~\ref{fig:Mn2P2S3Se3 orientation dependent band}). This can be understood as a competition between the out-of-plane magnetization and Rashba SOC~\cite{Niu2010,Yang2011}, which are essential ingredients to achieve $C_1\neq 0$, and the in-plane exchange interaction which disfavors such a topological phase~\cite{Kartsev2020}.
\begin{figure}[]
    \centering
    \includegraphics[width=\columnwidth]{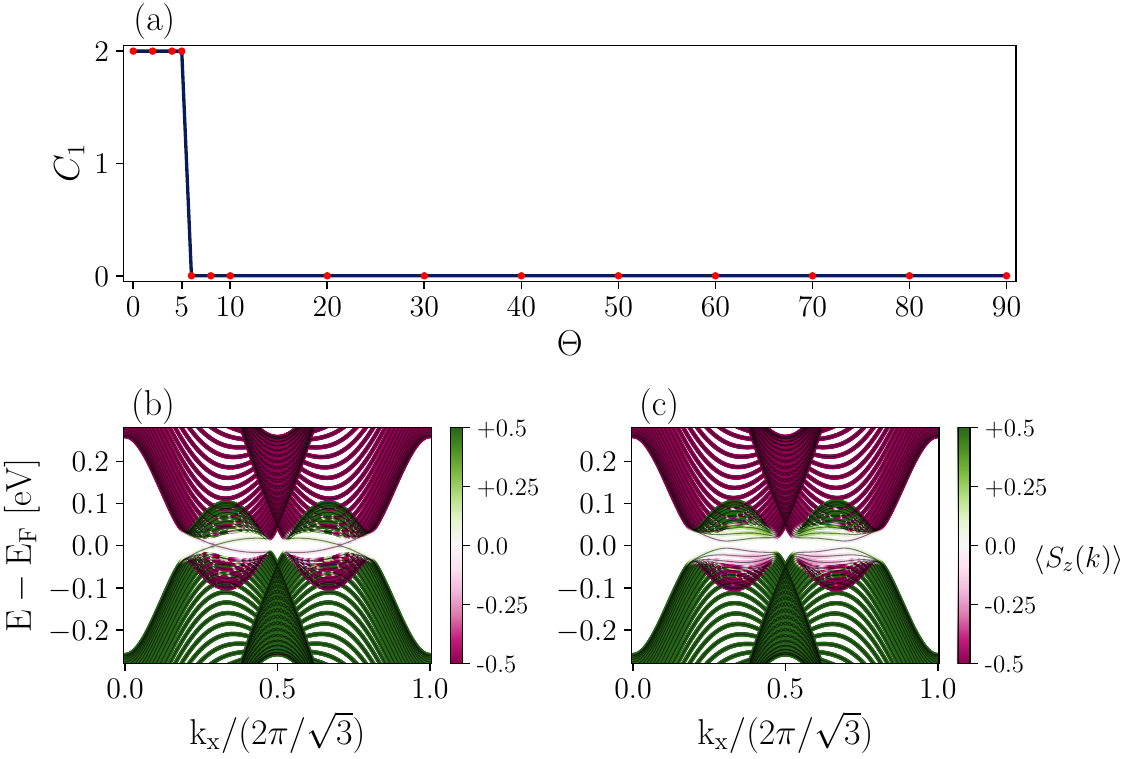}
    \caption{Tight-binding calculations of the evolution of the band topology of Mn$_2$P$_2$S$_3$Se$_3$ for different Mn spin orientations. (a) Calculated Chern number, $C_1$, as a function of polar angle. We find a topological phase transition from $C_1=2$ to $C_1=0$ near the critical polar angle $\Theta_c$. The gap closure at the critical polar angle $\Theta_c$, has a value almost the same as $\Theta_c\cong5.5^{\circ}$ obtained from DFT calculations. (b) Electronic structure of zigzag nanoribbon for $\Theta=0^{\circ}$. Two topological chiral edge modes associated with $C_1=2$ are present due to the bulk-edge correspondence.  (c) Electronic structure of zigzag nanoribbon for $\Theta=6^{\circ}$. At $\Theta=6^{\circ}$ after the phase transition, two gapped edge modes represent the trivial bulk topology ($C_1=0$).}
    \label{fig:Mn2P2S3Se3 TB_OBC}
\end{figure}  
\subsubsection{\label{sec:Effective TB of MnPS3Se3}Tight-binding Hamiltonian of Mn$_2$P$_2$S$_3$Se$_3$}
To explore possibilities of further tuning the topological nature of this compound, we construct the effective tight-binding Hamiltonian in Eq.~\eqref{eq:TB_Hamiltonian_MnPSe3} with nonzero Rashba SOC strength $\lambda_R$ that correctly captures characteristics of bands near the Fermi level and the topological phase transition at the critical polar angle $\Theta_c$ of $\mathit{\mathbf{\hat{m}}}$. Band structures are calculated from the tight-binding Hamiltonian at $\lambda_R/\lambda_{SO}=0.25,0.5$, and $0.75$ and shown in Fig.~\ref{fig:Mn2P2S3Se3 Rashba SOC strength}(a),(b) and (c). As observed in Fig.~\ref{fig:Mn2P2S3Se3 Rashba SOC strength}, the gap size increases with increasing $\lambda_R$, indicating that chemical substitutions in the isostructural Janus compounds that induce larger Rashba SOC strength than the (S, Se) pair, could be explored to enlarge the band gap. This approach would allow us to extend the energy range of the quantum anomalous Hall effect and achieve the Hall plateau in a larger energy window by circumventing the small electron pockets in this compound. 

In Fig.~\ref{fig:Mn2P2S3Se3 TB_OBC}(a), we observe the topological phase transition from $C_1=2$ to $C_1=0$ at the critical angle $\Theta_c$. The obtained value of $\Theta_c$ from the tight-binding calculations is consistent with the value obtained from DFT calculations, approximately $\Theta_c\cong5.5^{\circ}$. The topological nature of the system can be further confirmed by examining the chiral edge modes in a zigzag nanoribbon. In Fig.~\ref{fig:Mn2P2S3Se3 TB_OBC}(b), two topologically protected chiral edge modes are illustrated for $\Theta=0^{\circ}$. After the phase transition, these edge modes become gapped at $\Theta=6^{\circ}$, as shown in Fig.~\ref{fig:Mn2P2S3Se3 TB_OBC}(c). 

\section{\label{sec:conc}Conclusion}
We performed a comprehensive study of the electronic properties of the ferromagnetic monolayer \MnPSe~to understand the topological phases inherent to the band manifolds (band topology) in the presence of different magnetic space groups. In the FM$_z$ configuration of \MnPSe, we identify an intriguing accidental SM phase that is protected by both the order-two antiunitary composite symmetry $A=P\times(T\times M_y)$ and three-fold rotation symmetry $C_{3z}$. The topologically stable band crossing along $\Gamma$ to $\rm K$ can be achieved due to the Hamiltonian commuting with $A$ along $k_y=0$ , along with two other band crossings that are connected through the $C_{3z}$ symmetry. By manipulating the magnetic orientation of the Mn atoms, and thus breaking $C_{3z}$ symmetry with a non-zero polar angle $\Theta$, we are able to open a band gap. This result is confirmed by the band structures obtained from DFT calculations and the $\mathbf{k}\cdot\mathbf{p}$ Hamiltonian. Notably, the band gap reaches values of approximately $0.1$ eV when the magnetization is in the in-plane direction, suggesting the potential of utilizing this metal-to-insulator transition for switching and logic devices at room temperature just by manipulating the magnetization direction.

We present several symmetry designs and systematically study the interplay between magnetic space groups and strong first-order topological phases of the band manifolds below the gap at the Fermi energy. By breaking the inversion symmetry $P$ in \MnPSe, we observe the emergence of the topologically nontrivial bulk gap with the $C_1=-1$ invariant. To incorporate such symmetry breaking in a more realistic scenario, we investigate the isostructural Janus compound Mn$_2$P$_2$S$_3$Se$_3$ that naturally breaks $P$ symmetry. By tuning the polar angle of its magnetic orientation, we observe the topological phase transition from $C_1=2$ to $C_1=0$ as the gap closes at the critical polar angle $\Theta_c$. 

We construct an effective tight-binding Hamiltonian that accurately captures the band characteristics near the Fermi energy and reproduces the band topology and the topological phase transition obtained from DFT calculations. Such features are the result of the competition between exchange interaction and Rashba SOC, the latter appearing solely in the Janus compound due to the inversion symmetry breaking. Additionally, we find that by increasing the strength of the Rashba SOC, larger band gaps can be achieved. This suggests the possibility of exploring other Janus compounds with greater Rashba SOC strength than the (S,Se) pair in Mn$_2$P$_2$S$_3$Se$_3$, or applying perpendicular electric fields to also enhance the Rashba SOC \cite{Gmitra2009}, which could eliminate the small electron pockets and enable the realization of the quantum anomalous Hall phase with the Hall plateau in a larger energy window. 

In future research, it would be interesting to explore optical engineering as a means to access the hypothetical metastable FM \MnPSe~monolayer as the desired magnetic ground state. This involves harnessing the coherent light-matter interaction through tailored light pulses to drive the material out of equilibrium and manipulate its properties~\cite{Basov2017,Takashi2019,McIver2020,Torre2021}. By controlling the magnetic exchange interactions~\cite{Ron2020,Sriram2022}, this approach offers the potential to selectively induce magnetic phases that are not readily attainable under equilibrium conditions.

\begin{acknowledgements}
I. N. is grateful to Robert-Jan Slager for critical reading of our manuscript. M. V. is grateful to Luis M. Canonico for useful discussions and to T. V. Trevisan for help with the Qsymm package. This work is supported by the US Department of Energy, Office of Science, National Quantum Information Science Research Centers, Quantum Systems Accelerator (QSA). Computational resources were provided by the National Energy Research Scientific Computing Center and the Molecular Foundry, DOE Office of Science User Facilities supported under Contract No. DEAC02-05-CH11231. M.V. is supported as part of the Center for Novel Pathways to Quantum Coherence in Materials, an Energy Frontier Research Center funded by the US Department of Energy, Office of Science, Basic Energy Sciences. 
\end{acknowledgements}

\appendix

\section{\label{app:DFT details}Details of first-principles calculations and Wannierization}
In our calculations, the \MnPSe~monolayer is modeled using a periodic slab geometry in a supercell setup. To avoid spurious interactions with periodic images, a vacuum region of 28{\AA} in the out-of-plane direction is used. We employ a plane-wave basis with an energy cutoff of $520$ eV and use a dense k-point grid sampling of $13\times13\times1$ in the BZ according to the Monkhorst-Pack scheme~\cite{Monkhorst1976}. Both the lattice constant and positions of all atoms are relaxed until the force is less than $2$ meV/{\AA}, while keeping the volume fixed. The optimized lattice constant is found to be $a=b=6.363$ {\AA}, consistent with experimental value ($a=b=6.394$ {\AA})~\cite{Susner2017}. For the electronic self-consistent loop, the total energy convergence criterion is set to $10^{-7}$eV by using GGA (without U). This choice has shown consistent results when comparing calculations with different values of U, as reported in~\cite{Natalya2023}.

To calculate the topological invariants of the electronic band structures, we use the Wannier90 code and obtain the Wannier tight-binding Hamiltonian based on maximally localized Wannier functions (MLWFs)~\cite{Souza2001,Marzari2012}. To perform the Wannierization procedure, we carry out DFT calculations using a dense k-point grid of $18\times18\times1$. The choice of basis set for the Wannierization includes $p_{x,y,z}$ orbitals for the six Se atoms and $d_{xz,yz}$ orbitals for two Mn atoms present in the supercell. This selection allows us to construct a Wannier tight-binding Hamiltonian of rank $44$. Once we have the Wannier tight-binding Hamiltonian, we explicitly calculate the Berry curvature and the first Chern number~\cite{Fukui2005,Wang2007} from this Hamiltonian.  

\section{\label{app:TB details}Details of tight-binding model}
We construct the effective tight-binding Hamiltonian for the monolayer honeycomb lattice in Eq.~\eqref{eq:TB_Hamiltonian_MnPSe3}. We consider two spinful $d_{xz,yz}$ orbitals associated with the Mn atoms at each lattice site \cite{Sugita2018, Gu2019}. 

\subsection{\label{sec:nonmagnetic symmetries}Nonmagnetic symmetries in the lattice}

\begin{figure}[t]
    \centering
    \includegraphics[width=\columnwidth]{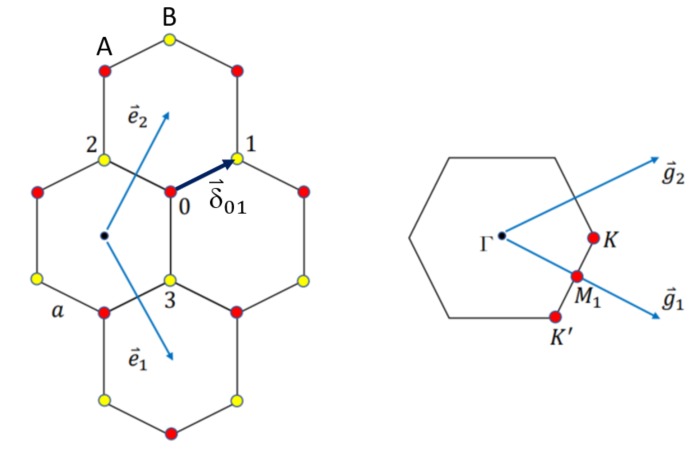}
    \caption{Lattice ($\mathbf{e}_{1,2}$) and reciprocal lattice ($\mathbf{g}_{1,2}$) basis vectors. A and B are the sublattices. $\mathbf{\delta}_{0,1}$ indicates the vector between nearest neighbor sites from A to B$_1$.}
    \label{fig:TB_Lattice}
\end{figure}

In Fig.~\ref{fig:TB_Lattice}, we choose the lattice basis vectors as
\begin{equation} \label{eq:lattice basis vectors}
    \mathbf{e}_1=\frac{\sqrt{3}}{2}\hat{x}+\frac{-3}{2}\hat{y},~ \mathbf{e}_2=\frac{\sqrt{3}}{2}\hat{x}+\frac{3}{2}\hat{y},
\end{equation}
by setting the bond length between two nearest-neighbor (NN) sites as $a=1$, the reciprocal lattice vectors are defined as $\mathbf{g}_1=2\pi/3\cdot(\sqrt{3}\hat{k}_x-\hat{k}_y)$ and $\mathbf{g}_2=2\pi/3\cdot(\sqrt{3}\hat{k}_x+\hat{k}_y)$, satisfying the condition $\mathbf{e}_i\cdot\mathbf{g}_j=2\pi\delta_{ij}$. The sites on the A(B) sublattice are located at positions $\mathbf{R}+\mathbf{t}_{A(B)}$, where $\mathbf{R}$ represents a lattice translation. Specifically, the sublattice vectors are given by 
\begin{equation} \label{eq:sublattice vectors}
    \mathbf{t}_{A}=\frac{2}{3}\mathbf{e}_1+\frac{1}{3}\mathbf{e}_2,~\mathbf{t}_{B}=\frac{1}{3}\mathbf{e}_1+\frac{2}{3}\mathbf{e}_2.
\end{equation}
In this basis, the symmetry generators of the honeycomb lattice itself (without considering magnetic moments on Mn), act as follows: 
\begin{align} \label{eq:symmetry generation}
    C_{3z}:&~(\mathbf{e}_1,\mathbf{e}_2)\rightarrow(\mathbf{e}_2,-\mathbf{e}_1-\mathbf{e}_2), \\
    P:&~(\mathbf{e}_1,\mathbf{e}_2)\rightarrow(-\mathbf{e}_1,-\mathbf{e}_2), \nonumber \\
    M_y:&~(\mathbf{e}_1,\mathbf{e}_2)\rightarrow(\mathbf{e}_2,\mathbf{e}_1), \nonumber
\end{align}
where $M_x$ is not preserved due to the coordination of six Se atoms. In the momentum representation of the Hamiltonian, the Pauli matrices $\sigma$, $\tau$, and $s$ act in sublattice, orbital, and spin space, respectively. We choose the following matrix representation for these matrices:
\begin{align} \label{eq:symmetry representation}
    U_{C_{3z}}=&~\sigma_0 \otimes e^{i2\pi/3\tau_y} \otimes e^{i\pi/3s_z}, \\
    U_{P}=&~\sigma_x \otimes \tau_0 \otimes s_0, \nonumber \\
    U_{M_y}:&~\sigma_x \otimes \tau_z \otimes e^{i\pi/2s_y}, \nonumber
\end{align}
satisfying $U_{C_{3z}}^3=-\mathbbm{1}$, $U_{P}^2=\mathbbm{1}$ and $U_{M_y}^2=-\mathbbm{1}$. In addition, we consider an order-two antiunitary composite symmetry of time-reversal and mirror symmetry, denoted as $T\times M_y$. It can be represented as $U_cK$ where 
\begin{equation}
    U_c=~\sigma_x \otimes \tau_z \otimes s_0  
\end{equation}
and $K$ represents complex conjugation. 

\subsection{\label{sec:hopping matrix imposed by symmetries}Symmetry restrictions on hopping matrices}
$c_{i,a,\alpha}$ in the Hamiltonian in Eq.~\eqref{eq:TB_Hamiltonian_MnPSe3} denotes the annihilation operator on site $i$, where $a=x,y$ indicates the $d_{xz}$ or $d_{yz}$ orbital, and $\alpha=x,y,z$ indicates the spin$-1/2$ degree of freedom. The hopping matrix $(t_{i,j})_{a,b}$ for $i=\mathbf{R}+\mathbf{t}_{A}$ and $j=\mathbf{R}+\mathbf{t}_{A}+\delta_{0r}$, where the $\delta_{0r}$ (with $r=1,2,3$) corresponds to each nearest B site from A site in $\mathbf{R}$, is expressed as the matrix $t_{0,r}=t_{r,0}^{\dag}$ given by
\begin{equation} \label{eq:hopping matrix}
    t_{0,1}=\begin{bmatrix}
            t_{x,x} & t_{x,y} \\ 
            t_{y,x} & t_{y,y}, 
\end{bmatrix}
\end{equation}  
where each $(t_{0,1})_{a,b}=(t_{0,1})^{R}_{a,b}+i(t_{0,1})^{Im}_{a,b}$ is a complex number without time-reversal symmetry (imposing each element of the matrix $t_{0,1}$ to be real). The symmetries $C_{3z}$ and $M_y$ in Eq.~\eqref{eq:symmetry representation} impose the following constraints on the hopping matrix elements: 
\begin{align} \label{eq:symmetry restriction on hopping matrix}
    t_{2,0}=&t_{0,2}^{\dag}=U_{M_y}t_{0,1}U_{M_y}^{\dag}, \\
    t_{0,2}=&U_{C_{3z}}t_{0,1}U_{C_{3z}}^{\dag}, \nonumber \\ t_{0,3}=&U_{C_{3z}}^{\dag}t_{0,1}U_{C_{3z}}. \nonumber
\end{align}
From Eq.~\eqref{eq:symmetry restriction on hopping matrix}, the hopping elements of $t_{0,1}$ need to satisfy the following condition: $\operatorname{Im}(t_{x,x})=-\operatorname{Im}(t_{y,y})$, $\sqrt{3}\cdot(t_{x,y}^{R}-t_{y,y}^{R})=(t_{x,y}^{R}+t_{y,x}^{R})$, and $\operatorname{Im}(t_{x,x})=-\sqrt{3}\operatorname{Im}(t_{x,y})$. These conditions result in four independent parameters: $t_{x,x}^{R},t_{x,x}^{Im},t_{x,y}^{R},$ and $t_{y,y}^{R}$. In order to preserve the $P$ inversion symmetry, an additional constraint $t_{0,1}=t_{0,1}^{\dag}$ needs to be satisfied, which implies $\operatorname{Im}(t_{x,x})=0$ and $t_{x,y}^{R}=t_{y,x}^{R}=\sqrt{3}/2\cdot(t_{x,x}^{R}-t_{y,y}^{R})$. As a result, the number of independent parameters reduces to two: $t_{x,x}^{R}$ and $t_{y,y}^{R}$. 

Additionally, for the composite symmetry $T\times M_y$ to be preserved, the following condition needs to be satisfied: 
\begin{equation} \label{eq:composite symmetry restriction on hopping matrix}
   t_{2,0}=t_{0,2}^{\dag}=\tau_zt_{0,1}^{*}\tau_z=e^{-i2\pi/3\tau_y}t_{0,1}^{\dag}e^{i2\pi/3\tau_y}. 
\end{equation}
In the case where the system preserves $P$, $C_{3z}$, and $M_y$ symmetries, this condition~\eqref{eq:composite symmetry restriction on hopping matrix} is automatically satisfied. However, when only $C_{3z}$ and $M_y$ are preserved (breaking $P$), an additional condition $\operatorname{Im}(t_{x,x})=0$ is imposed. Therefore, in this case, only three independent parameters $t_{x,x}^{R}$, $t_{x,y}^{R}$, and $t_{y,y}^{R}$ are needed. 

\begin{table}[h]
\caption{\label{tab:Symmetry table of the tight-binding Hamiltonian} Symmetries of different terms in the tight-binding Hamiltonian $H=H_{hop}+H_{\lambda_{SO}}+H_{\lambda_R}+H_M$ in Eq.~\eqref{eq:TB_Hamiltonian_MnPSe3}.} 
\begin{ruledtabular}
\centering
\begin{tabular}{|| c | c | c | l | l | l ||} 
  Symmetry & $H_{\lambda_{SO}} (\textrm{SOC})$ & $H_{\lambda_R} (\textrm{SOC})$ & $\mathbf{\hat{m}}_x$ & $\mathbf{\hat{m}}_y$ & $\mathbf{\hat{m}}_z$ \\
 \hline\hline
 $C_{3z}$ & $\textcolor{teal}{\checkmark}$ & $\textcolor{teal}{\checkmark}$ & $\times$ & $\times$ & $\textcolor{teal}{\checkmark}$ \\ 
 \hline
 $P$ & $\textcolor{teal}{\checkmark}$ & $\times$ & $\textcolor{teal}{\checkmark}$ & $\textcolor{teal}{\checkmark}$ & $\textcolor{teal}{\checkmark}$ \\
 \hline
 $M_y$ & $\textcolor{teal}{\checkmark}$ & $\textcolor{teal}{\checkmark}$ & $\times$ & $\textcolor{teal}{\checkmark}$ & $\times$ \\
 \hline
 $T\times M_y$ & $\textcolor{teal}{\checkmark}$ & $\textcolor{teal}{\checkmark}$ & $\textcolor{teal}{\checkmark}$ & $\times$ & $\textcolor{teal}{\checkmark}$ \\
\end{tabular}
\end{ruledtabular}
\end{table}

\subsection{\label{sec:existing symmetries in magnetization}Existing symmetries in tight-binding model}
The Hamiltonian in Eq.~\eqref{eq:TB_Hamiltonian_MnPSe3} can be decomposed into different terms: $H=H_{hop}+H_{\lambda_{SO}}+H_{\lambda_R}+H_{M}$. $H_{\lambda_{SO}}$ represents the intrinsic SOC Hamiltonian, while $H_{\lambda_{R}}$ represents the extrinsic Rashba SOC Hamiltonian. $H_{M}$ represents the uniform exchange Zeeman term determined from the magnetic orientation of the ferromagnetic (FM) Mn. We identify the NN hopping matrices in $H_{hop}$ that preserve $C_{3z}$ and $M_y$ symmetries among nonmagnetic symmetries of the lattice given in Eq.~\eqref{eq:symmetry generation}. We further classify these hopping matrices into two additional cases: $(1)$ the existence of $P$ (inversion symmetry) and $(2)$ the existence of $T\times M_y$ (composite symmetry of time-reversal and mirror symmetry). 

In Table~\ref{tab:Symmetry table of the tight-binding Hamiltonian}, we analyze the existing symmetries in the tight-binding Hamiltonian in Eq.~\eqref{eq:TB_Hamiltonian_MnPSe3}. Nonzero Rashba SOC breaks the $P$ symmetry. The presence of nonzero magnetization explicitly breaks the $T$ symmetry, and the orientation of the magnetization vector $\mathit{\mathbf{\hat{m}}}$ selectively breaks the crystalline symmetries. The composite symmetry $T\times M_y$ is preserved unless $\mathbf{\hat{m}}_y\neq0$, and any in-plane components of $\mathit{\mathbf{\hat{m}}}$ break $C_{3z}$ symmetry.  

\bibliography{refs}

\begin{thebibliography}{76}%
\makeatletter
\providecommand \@ifxundefined [1]{%
 \@ifx{#1\undefined}
}%
\providecommand \@ifnum [1]{%
 \ifnum #1\expandafter \@firstoftwo
 \else \expandafter \@secondoftwo
 \fi
}%
\providecommand \@ifx [1]{%
 \ifx #1\expandafter \@firstoftwo
 \else \expandafter \@secondoftwo
 \fi
}%
\providecommand \natexlab [1]{#1}%
\providecommand \enquote  [1]{``#1''}%
\providecommand \bibnamefont  [1]{#1}%
\providecommand \bibfnamefont [1]{#1}%
\providecommand \citenamefont [1]{#1}%
\providecommand \href@noop [0]{\@secondoftwo}%
\providecommand \href [0]{\begingroup \@sanitize@url \@href}%
\providecommand \@href[1]{\@@startlink{#1}\@@href}%
\providecommand \@@href[1]{\endgroup#1\@@endlink}%
\providecommand \@sanitize@url [0]{\catcode `\\12\catcode `\$12\catcode
  `\&12\catcode `\#12\catcode `\^12\catcode `\_12\catcode `\%12\relax}%
\providecommand \@@startlink[1]{}%
\providecommand \@@endlink[0]{}%
\providecommand \url  [0]{\begingroup\@sanitize@url \@url }%
\providecommand \@url [1]{\endgroup\@href {#1}{\urlprefix }}%
\providecommand \urlprefix  [0]{URL }%
\providecommand \Eprint [0]{\href }%
\providecommand \doibase [0]{https://doi.org/}%
\providecommand \selectlanguage [0]{\@gobble}%
\providecommand \bibinfo  [0]{\@secondoftwo}%
\providecommand \bibfield  [0]{\@secondoftwo}%
\providecommand \translation [1]{[#1]}%
\providecommand \BibitemOpen [0]{}%
\providecommand \bibitemStop [0]{}%
\providecommand \bibitemNoStop [0]{.\EOS\space}%
\providecommand \EOS [0]{\spacefactor3000\relax}%
\providecommand \BibitemShut  [1]{\csname bibitem#1\endcsname}%
\let\auto@bib@innerbib\@empty
\bibitem [{\citenamefont {Frey}\ \emph {et~al.}(2020)\citenamefont {Frey},
  \citenamefont {Horton}, \citenamefont {Munro}, \citenamefont {Griffin},
  \citenamefont {Persson},\ and\ \citenamefont {Shenoy}}]{Frey2020}%
  \BibitemOpen
  \bibfield  {author} {\bibinfo {author} {\bibfnamefont {N.~C.}\ \bibnamefont
  {Frey}}, \bibinfo {author} {\bibfnamefont {M.~K.}\ \bibnamefont {Horton}},
  \bibinfo {author} {\bibfnamefont {J.~M.}\ \bibnamefont {Munro}}, \bibinfo
  {author} {\bibfnamefont {S.~M.}\ \bibnamefont {Griffin}}, \bibinfo {author}
  {\bibfnamefont {K.~A.}\ \bibnamefont {Persson}},\ and\ \bibinfo {author}
  {\bibfnamefont {V.~B.}\ \bibnamefont {Shenoy}},\ }\bibfield  {title}
  {\bibinfo {title} {High-throughput search for magnetic and topological order
  in transition metal oxides},\ }\href {https://doi.org/10.1126/sciadv.abd1076}
  {\bibfield  {journal} {\bibinfo  {journal} {Science Advances}\ }\textbf
  {\bibinfo {volume} {6}},\ \bibinfo {pages} {eabd1076} (\bibinfo {year}
  {2020})},\ \Eprint
  {https://arxiv.org/abs/https://www.science.org/doi/pdf/10.1126/sciadv.abd1076}
  {https://www.science.org/doi/pdf/10.1126/sciadv.abd1076} \BibitemShut
  {NoStop}%
\bibitem [{\citenamefont {Xu}\ \emph {et~al.}(2020)\citenamefont {Xu},
  \citenamefont {Elcoro}, \citenamefont {Song}, \citenamefont {Wieder},
  \citenamefont {Vergniory}, \citenamefont {Regnault}, \citenamefont {Chen},
  \citenamefont {Felser},\ and\ \citenamefont {Bernevig}}]{Xu2020}%
  \BibitemOpen
  \bibfield  {author} {\bibinfo {author} {\bibfnamefont {Y.}~\bibnamefont
  {Xu}}, \bibinfo {author} {\bibfnamefont {L.}~\bibnamefont {Elcoro}}, \bibinfo
  {author} {\bibfnamefont {Z.-D.}\ \bibnamefont {Song}}, \bibinfo {author}
  {\bibfnamefont {B.~J.}\ \bibnamefont {Wieder}}, \bibinfo {author}
  {\bibfnamefont {M.~G.}\ \bibnamefont {Vergniory}}, \bibinfo {author}
  {\bibfnamefont {N.}~\bibnamefont {Regnault}}, \bibinfo {author}
  {\bibfnamefont {Y.}~\bibnamefont {Chen}}, \bibinfo {author} {\bibfnamefont
  {C.}~\bibnamefont {Felser}},\ and\ \bibinfo {author} {\bibfnamefont {B.~A.}\
  \bibnamefont {Bernevig}},\ }\bibfield  {title} {\bibinfo {title}
  {High-throughput calculations of magnetic topological materials},\ }\href
  {https://doi.org/10.1038/s41586-020-2837-0} {\bibfield  {journal} {\bibinfo
  {journal} {Nature}\ }\textbf {\bibinfo {volume} {586}},\ \bibinfo {pages}
  {702} (\bibinfo {year} {2020})}\BibitemShut {NoStop}%
\bibitem [{\citenamefont {Elcoro}\ \emph {et~al.}(2021)\citenamefont {Elcoro},
  \citenamefont {Wieder}, \citenamefont {Song}, \citenamefont {Xu},
  \citenamefont {Bradlyn},\ and\ \citenamefont {Bernevig}}]{Elcoro2021}%
  \BibitemOpen
  \bibfield  {author} {\bibinfo {author} {\bibfnamefont {L.}~\bibnamefont
  {Elcoro}}, \bibinfo {author} {\bibfnamefont {B.~J.}\ \bibnamefont {Wieder}},
  \bibinfo {author} {\bibfnamefont {Z.}~\bibnamefont {Song}}, \bibinfo {author}
  {\bibfnamefont {Y.}~\bibnamefont {Xu}}, \bibinfo {author} {\bibfnamefont
  {B.}~\bibnamefont {Bradlyn}},\ and\ \bibinfo {author} {\bibfnamefont {B.~A.}\
  \bibnamefont {Bernevig}},\ }\bibfield  {title} {\bibinfo {title} {Magnetic
  topological quantum chemistry},\ }\href
  {https://doi.org/10.1038/s41467-021-26241-8} {\bibfield  {journal} {\bibinfo
  {journal} {Nature Communications}\ }\textbf {\bibinfo {volume} {12}},\
  \bibinfo {pages} {5965} (\bibinfo {year} {2021})}\BibitemShut {NoStop}%
\bibitem [{\citenamefont {Bernevig}\ \emph {et~al.}(2022)\citenamefont
  {Bernevig}, \citenamefont {Felser},\ and\ \citenamefont
  {Beidenkopf}}]{Bernevig2022}%
  \BibitemOpen
  \bibfield  {author} {\bibinfo {author} {\bibfnamefont {B.~A.}\ \bibnamefont
  {Bernevig}}, \bibinfo {author} {\bibfnamefont {C.}~\bibnamefont {Felser}},\
  and\ \bibinfo {author} {\bibfnamefont {H.}~\bibnamefont {Beidenkopf}},\
  }\bibfield  {title} {\bibinfo {title} {Progress and prospects in magnetic
  topological materials},\ }\href {https://doi.org/10.1038/s41586-021-04105-x}
  {\bibfield  {journal} {\bibinfo  {journal} {Nature}\ }\textbf {\bibinfo
  {volume} {603}},\ \bibinfo {pages} {41} (\bibinfo {year} {2022})}\BibitemShut
  {NoStop}%
\bibitem [{\citenamefont {Gong}\ \emph {et~al.}(2017)\citenamefont {Gong},
  \citenamefont {Li}, \citenamefont {Li}, \citenamefont {Ji}, \citenamefont
  {Stern}, \citenamefont {Xia}, \citenamefont {Cao}, \citenamefont {Bao},
  \citenamefont {Wang}, \citenamefont {Wang}, \citenamefont {Qiu},
  \citenamefont {Cava}, \citenamefont {Louie}, \citenamefont {Xia},\ and\
  \citenamefont {Zhang}}]{Gong2017}%
  \BibitemOpen
  \bibfield  {author} {\bibinfo {author} {\bibfnamefont {C.}~\bibnamefont
  {Gong}}, \bibinfo {author} {\bibfnamefont {L.}~\bibnamefont {Li}}, \bibinfo
  {author} {\bibfnamefont {Z.}~\bibnamefont {Li}}, \bibinfo {author}
  {\bibfnamefont {H.}~\bibnamefont {Ji}}, \bibinfo {author} {\bibfnamefont
  {A.}~\bibnamefont {Stern}}, \bibinfo {author} {\bibfnamefont
  {Y.}~\bibnamefont {Xia}}, \bibinfo {author} {\bibfnamefont {T.}~\bibnamefont
  {Cao}}, \bibinfo {author} {\bibfnamefont {W.}~\bibnamefont {Bao}}, \bibinfo
  {author} {\bibfnamefont {C.}~\bibnamefont {Wang}}, \bibinfo {author}
  {\bibfnamefont {Y.}~\bibnamefont {Wang}}, \bibinfo {author} {\bibfnamefont
  {Z.~Q.}\ \bibnamefont {Qiu}}, \bibinfo {author} {\bibfnamefont {R.~J.}\
  \bibnamefont {Cava}}, \bibinfo {author} {\bibfnamefont {S.~G.}\ \bibnamefont
  {Louie}}, \bibinfo {author} {\bibfnamefont {J.}~\bibnamefont {Xia}},\ and\
  \bibinfo {author} {\bibfnamefont {X.}~\bibnamefont {Zhang}},\ }\bibfield
  {title} {\bibinfo {title} {Discovery of intrinsic ferromagnetism in
  two-dimensional van der {W}aals crystals},\ }\href
  {https://doi.org/10.1038/nature22060} {\bibfield  {journal} {\bibinfo
  {journal} {Nature}\ }\textbf {\bibinfo {volume} {546}},\ \bibinfo {pages}
  {265} (\bibinfo {year} {2017})}\BibitemShut {NoStop}%
\bibitem [{\citenamefont {Huang}\ \emph {et~al.}(2017)\citenamefont {Huang},
  \citenamefont {Clark}, \citenamefont {Navarro-Moratalla}, \citenamefont
  {Klein}, \citenamefont {Cheng}, \citenamefont {Seyler}, \citenamefont
  {Zhong}, \citenamefont {Schmidgall}, \citenamefont {McGuire}, \citenamefont
  {Cobden}, \citenamefont {Yao}, \citenamefont {Xiao}, \citenamefont
  {Jarillo-Herrero},\ and\ \citenamefont {Xu}}]{Huang2017}%
  \BibitemOpen
  \bibfield  {author} {\bibinfo {author} {\bibfnamefont {B.}~\bibnamefont
  {Huang}}, \bibinfo {author} {\bibfnamefont {G.}~\bibnamefont {Clark}},
  \bibinfo {author} {\bibfnamefont {E.}~\bibnamefont {Navarro-Moratalla}},
  \bibinfo {author} {\bibfnamefont {D.~R.}\ \bibnamefont {Klein}}, \bibinfo
  {author} {\bibfnamefont {R.}~\bibnamefont {Cheng}}, \bibinfo {author}
  {\bibfnamefont {K.~L.}\ \bibnamefont {Seyler}}, \bibinfo {author}
  {\bibfnamefont {D.}~\bibnamefont {Zhong}}, \bibinfo {author} {\bibfnamefont
  {E.}~\bibnamefont {Schmidgall}}, \bibinfo {author} {\bibfnamefont {M.~A.}\
  \bibnamefont {McGuire}}, \bibinfo {author} {\bibfnamefont {D.~H.}\
  \bibnamefont {Cobden}}, \bibinfo {author} {\bibfnamefont {W.}~\bibnamefont
  {Yao}}, \bibinfo {author} {\bibfnamefont {D.}~\bibnamefont {Xiao}}, \bibinfo
  {author} {\bibfnamefont {P.}~\bibnamefont {Jarillo-Herrero}},\ and\ \bibinfo
  {author} {\bibfnamefont {X.}~\bibnamefont {Xu}},\ }\bibfield  {title}
  {\bibinfo {title} {Layer-dependent ferromagnetism in a van der {W}aals
  crystal down to the monolayer limit},\ }\href
  {https://doi.org/10.1038/nature22391} {\bibfield  {journal} {\bibinfo
  {journal} {Nature}\ }\textbf {\bibinfo {volume} {546}},\ \bibinfo {pages}
  {270} (\bibinfo {year} {2017})}\BibitemShut {NoStop}%
\bibitem [{\citenamefont {Wang}\ \emph {et~al.}(2013)\citenamefont {Wang},
  \citenamefont {Alzate},\ and\ \citenamefont {Amiri}}]{wang2013low}%
  \BibitemOpen
  \bibfield  {author} {\bibinfo {author} {\bibfnamefont {K.}~\bibnamefont
  {Wang}}, \bibinfo {author} {\bibfnamefont {J.}~\bibnamefont {Alzate}},\ and\
  \bibinfo {author} {\bibfnamefont {P.~K.}\ \bibnamefont {Amiri}},\ }\bibfield
  {title} {\bibinfo {title} {Low-power non-volatile spintronic memory:
  {STT-RAM} and beyond},\ }\href@noop {} {\bibfield  {journal} {\bibinfo
  {journal} {Journal of Physics D: Applied Physics}\ }\textbf {\bibinfo
  {volume} {46}},\ \bibinfo {pages} {074003} (\bibinfo {year}
  {2013})}\BibitemShut {NoStop}%
\bibitem [{\citenamefont {Kurebayashi}\ \emph {et~al.}(2022)\citenamefont
  {Kurebayashi}, \citenamefont {Garcia}, \citenamefont {Khan}, \citenamefont
  {Sinova},\ and\ \citenamefont {Roche}}]{Kurebayashi2022}%
  \BibitemOpen
  \bibfield  {author} {\bibinfo {author} {\bibfnamefont {H.}~\bibnamefont
  {Kurebayashi}}, \bibinfo {author} {\bibfnamefont {J.~H.}\ \bibnamefont
  {Garcia}}, \bibinfo {author} {\bibfnamefont {S.}~\bibnamefont {Khan}},
  \bibinfo {author} {\bibfnamefont {J.}~\bibnamefont {Sinova}},\ and\ \bibinfo
  {author} {\bibfnamefont {S.}~\bibnamefont {Roche}},\ }\bibfield  {title}
  {\bibinfo {title} {Magnetism, symmetry and spin transport in van der {W}aals
  layered systems},\ }\href {https://doi.org/10.1038/s42254-021-00403-5}
  {\bibfield  {journal} {\bibinfo  {journal} {Nature Reviews Physics}\ }\textbf
  {\bibinfo {volume} {4}},\ \bibinfo {pages} {150} (\bibinfo {year}
  {2022})}\BibitemShut {NoStop}%
\bibitem [{\citenamefont {Yang}\ \emph {et~al.}(2022)\citenamefont {Yang},
  \citenamefont {Valenzuela}, \citenamefont {Chshiev}, \citenamefont {Couet},
  \citenamefont {Dieny}, \citenamefont {Dlubak}, \citenamefont {Fert},
  \citenamefont {Garello}, \citenamefont {Jamet}, \citenamefont {Jeong},
  \citenamefont {Lee}, \citenamefont {Lee}, \citenamefont {Martin},
  \citenamefont {Kar}, \citenamefont {S{\'e}n{\'e}or}, \citenamefont {Shin},\
  and\ \citenamefont {Roche}}]{Yang2022}%
  \BibitemOpen
  \bibfield  {author} {\bibinfo {author} {\bibfnamefont {H.}~\bibnamefont
  {Yang}}, \bibinfo {author} {\bibfnamefont {S.~O.}\ \bibnamefont
  {Valenzuela}}, \bibinfo {author} {\bibfnamefont {M.}~\bibnamefont {Chshiev}},
  \bibinfo {author} {\bibfnamefont {S.}~\bibnamefont {Couet}}, \bibinfo
  {author} {\bibfnamefont {B.}~\bibnamefont {Dieny}}, \bibinfo {author}
  {\bibfnamefont {B.}~\bibnamefont {Dlubak}}, \bibinfo {author} {\bibfnamefont
  {A.}~\bibnamefont {Fert}}, \bibinfo {author} {\bibfnamefont {K.}~\bibnamefont
  {Garello}}, \bibinfo {author} {\bibfnamefont {M.}~\bibnamefont {Jamet}},
  \bibinfo {author} {\bibfnamefont {D.-E.}\ \bibnamefont {Jeong}}, \bibinfo
  {author} {\bibfnamefont {K.}~\bibnamefont {Lee}}, \bibinfo {author}
  {\bibfnamefont {T.}~\bibnamefont {Lee}}, \bibinfo {author} {\bibfnamefont
  {M.-B.}\ \bibnamefont {Martin}}, \bibinfo {author} {\bibfnamefont {G.~S.}\
  \bibnamefont {Kar}}, \bibinfo {author} {\bibfnamefont {P.}~\bibnamefont
  {S{\'e}n{\'e}or}}, \bibinfo {author} {\bibfnamefont {H.-J.}\ \bibnamefont
  {Shin}},\ and\ \bibinfo {author} {\bibfnamefont {S.}~\bibnamefont {Roche}},\
  }\bibfield  {title} {\bibinfo {title} {Two-dimensional materials prospects
  for non-volatile spintronic memories},\ }\href
  {https://doi.org/10.1038/s41586-022-04768-0} {\bibfield  {journal} {\bibinfo
  {journal} {Nature}\ }\textbf {\bibinfo {volume} {606}},\ \bibinfo {pages}
  {663} (\bibinfo {year} {2022})}\BibitemShut {NoStop}%
\bibitem [{\citenamefont {Husremovi{\'{c}}}\ \emph {et~al.}(2022)\citenamefont
  {Husremovi{\'{c}}}, \citenamefont {Groschner}, \citenamefont {Inzani},
  \citenamefont {Craig}, \citenamefont {Bustillo}, \citenamefont {Ercius},
  \citenamefont {Kazmierczak}, \citenamefont {Syndikus}, \citenamefont
  {Van~Winkle}, \citenamefont {Aloni}, \citenamefont {Taniguchi}, \citenamefont
  {Watanabe}, \citenamefont {Griffin},\ and\ \citenamefont
  {Bediako}}]{Husremović2022}%
  \BibitemOpen
  \bibfield  {author} {\bibinfo {author} {\bibfnamefont {S.}~\bibnamefont
  {Husremovi{\'{c}}}}, \bibinfo {author} {\bibfnamefont {C.~K.}\ \bibnamefont
  {Groschner}}, \bibinfo {author} {\bibfnamefont {K.}~\bibnamefont {Inzani}},
  \bibinfo {author} {\bibfnamefont {I.~M.}\ \bibnamefont {Craig}}, \bibinfo
  {author} {\bibfnamefont {K.~C.}\ \bibnamefont {Bustillo}}, \bibinfo {author}
  {\bibfnamefont {P.}~\bibnamefont {Ercius}}, \bibinfo {author} {\bibfnamefont
  {N.~P.}\ \bibnamefont {Kazmierczak}}, \bibinfo {author} {\bibfnamefont
  {J.}~\bibnamefont {Syndikus}}, \bibinfo {author} {\bibfnamefont
  {M.}~\bibnamefont {Van~Winkle}}, \bibinfo {author} {\bibfnamefont
  {S.}~\bibnamefont {Aloni}}, \bibinfo {author} {\bibfnamefont
  {T.}~\bibnamefont {Taniguchi}}, \bibinfo {author} {\bibfnamefont
  {K.}~\bibnamefont {Watanabe}}, \bibinfo {author} {\bibfnamefont {S.~M.}\
  \bibnamefont {Griffin}},\ and\ \bibinfo {author} {\bibfnamefont {D.~K.}\
  \bibnamefont {Bediako}},\ }\bibfield  {title} {\bibinfo {title} {Hard
  ferromagnetism down to the thinnest limit of iron-intercalated tantalum
  disulfide},\ }\href {https://doi.org/10.1021/jacs.2c02885} {\bibfield
  {journal} {\bibinfo  {journal} {Journal of the American Chemical Society}\
  }\textbf {\bibinfo {volume} {144}},\ \bibinfo {pages} {12167} (\bibinfo
  {year} {2022})}\BibitemShut {NoStop}%
\bibitem [{\citenamefont {Wu}\ \emph {et~al.}(2020)\citenamefont {Wu},
  \citenamefont {Zhang}, \citenamefont {Zhang}, \citenamefont {Wang},
  \citenamefont {Zhu}, \citenamefont {Hu}, \citenamefont {Yin}, \citenamefont
  {Wong}, \citenamefont {Fang}, \citenamefont {Wan}, \citenamefont {Han},
  \citenamefont {Shao}, \citenamefont {Taniguchi}, \citenamefont {Watanabe},
  \citenamefont {Zang}, \citenamefont {Mao}, \citenamefont {Zhang},\ and\
  \citenamefont {Wang}}]{Wu2020}%
  \BibitemOpen
  \bibfield  {author} {\bibinfo {author} {\bibfnamefont {Y.}~\bibnamefont
  {Wu}}, \bibinfo {author} {\bibfnamefont {S.}~\bibnamefont {Zhang}}, \bibinfo
  {author} {\bibfnamefont {J.}~\bibnamefont {Zhang}}, \bibinfo {author}
  {\bibfnamefont {W.}~\bibnamefont {Wang}}, \bibinfo {author} {\bibfnamefont
  {Y.~L.}\ \bibnamefont {Zhu}}, \bibinfo {author} {\bibfnamefont
  {J.}~\bibnamefont {Hu}}, \bibinfo {author} {\bibfnamefont {G.}~\bibnamefont
  {Yin}}, \bibinfo {author} {\bibfnamefont {K.}~\bibnamefont {Wong}}, \bibinfo
  {author} {\bibfnamefont {C.}~\bibnamefont {Fang}}, \bibinfo {author}
  {\bibfnamefont {C.}~\bibnamefont {Wan}}, \bibinfo {author} {\bibfnamefont
  {X.}~\bibnamefont {Han}}, \bibinfo {author} {\bibfnamefont {Q.}~\bibnamefont
  {Shao}}, \bibinfo {author} {\bibfnamefont {T.}~\bibnamefont {Taniguchi}},
  \bibinfo {author} {\bibfnamefont {K.}~\bibnamefont {Watanabe}}, \bibinfo
  {author} {\bibfnamefont {J.}~\bibnamefont {Zang}}, \bibinfo {author}
  {\bibfnamefont {Z.}~\bibnamefont {Mao}}, \bibinfo {author} {\bibfnamefont
  {X.}~\bibnamefont {Zhang}},\ and\ \bibinfo {author} {\bibfnamefont {K.~L.}\
  \bibnamefont {Wang}},\ }\bibfield  {title} {\bibinfo {title} {N{\'e}el-type
  skyrmion in {WTe$_2$/Fe$_3$GeTe$_2$} van der {W}aals heterostructure},\
  }\href {https://doi.org/10.1038/s41467-020-17566-x} {\bibfield  {journal}
  {\bibinfo  {journal} {Nature Communications}\ }\textbf {\bibinfo {volume}
  {11}},\ \bibinfo {pages} {3860} (\bibinfo {year} {2020})}\BibitemShut
  {NoStop}%
\bibitem [{\citenamefont {Ding}\ \emph {et~al.}(2020)\citenamefont {Ding},
  \citenamefont {Li}, \citenamefont {Xu}, \citenamefont {Li}, \citenamefont
  {Hou}, \citenamefont {Liu}, \citenamefont {Xi}, \citenamefont {Xu},
  \citenamefont {Yao},\ and\ \citenamefont {Wang}}]{Ding2020}%
  \BibitemOpen
  \bibfield  {author} {\bibinfo {author} {\bibfnamefont {B.}~\bibnamefont
  {Ding}}, \bibinfo {author} {\bibfnamefont {Z.}~\bibnamefont {Li}}, \bibinfo
  {author} {\bibfnamefont {G.}~\bibnamefont {Xu}}, \bibinfo {author}
  {\bibfnamefont {H.}~\bibnamefont {Li}}, \bibinfo {author} {\bibfnamefont
  {Z.}~\bibnamefont {Hou}}, \bibinfo {author} {\bibfnamefont {E.}~\bibnamefont
  {Liu}}, \bibinfo {author} {\bibfnamefont {X.}~\bibnamefont {Xi}}, \bibinfo
  {author} {\bibfnamefont {F.}~\bibnamefont {Xu}}, \bibinfo {author}
  {\bibfnamefont {Y.}~\bibnamefont {Yao}},\ and\ \bibinfo {author}
  {\bibfnamefont {W.}~\bibnamefont {Wang}},\ }\bibfield  {title} {\bibinfo
  {title} {Observation of magnetic skyrmion bubbles in a van der {W}aals
  ferromagnet {Fe$_3$GeTe$_2$}},\ }\href
  {https://doi.org/10.1021/acs.nanolett.9b03453} {\bibfield  {journal}
  {\bibinfo  {journal} {Nano Letters}\ }\textbf {\bibinfo {volume} {20}},\
  \bibinfo {pages} {868} (\bibinfo {year} {2020})}\BibitemShut {NoStop}%
\bibitem [{\citenamefont {Li}\ \emph {et~al.}(2019)\citenamefont {Li},
  \citenamefont {Wu}, \citenamefont {Gu}, \citenamefont {Le}, \citenamefont
  {Qin}, \citenamefont {Thomale},\ and\ \citenamefont {Hu}}]{Li2019}%
  \BibitemOpen
  \bibfield  {author} {\bibinfo {author} {\bibfnamefont {Y.}~\bibnamefont
  {Li}}, \bibinfo {author} {\bibfnamefont {X.}~\bibnamefont {Wu}}, \bibinfo
  {author} {\bibfnamefont {Y.}~\bibnamefont {Gu}}, \bibinfo {author}
  {\bibfnamefont {C.}~\bibnamefont {Le}}, \bibinfo {author} {\bibfnamefont
  {S.}~\bibnamefont {Qin}}, \bibinfo {author} {\bibfnamefont {R.}~\bibnamefont
  {Thomale}},\ and\ \bibinfo {author} {\bibfnamefont {J.}~\bibnamefont {Hu}},\
  }\bibfield  {title} {\bibinfo {title} {Topological superconductivity in
  {Ni}-based transition metal trichalcogenides},\ }\href
  {https://doi.org/10.1103/PhysRevB.100.214513} {\bibfield  {journal} {\bibinfo
   {journal} {Phys. Rev. B}\ }\textbf {\bibinfo {volume} {100}},\ \bibinfo
  {pages} {214513} (\bibinfo {year} {2019})}\BibitemShut {NoStop}%
\bibitem [{\citenamefont {Kezilebieke}\ \emph {et~al.}(2020)\citenamefont
  {Kezilebieke}, \citenamefont {Huda}, \citenamefont {Va{\v{n}}o},
  \citenamefont {Aapro}, \citenamefont {Ganguli}, \citenamefont {Silveira},
  \citenamefont {G{\l}odzik}, \citenamefont {Foster}, \citenamefont {Ojanen},\
  and\ \citenamefont {Liljeroth}}]{Kezilebieke2020}%
  \BibitemOpen
  \bibfield  {author} {\bibinfo {author} {\bibfnamefont {S.}~\bibnamefont
  {Kezilebieke}}, \bibinfo {author} {\bibfnamefont {M.~N.}\ \bibnamefont
  {Huda}}, \bibinfo {author} {\bibfnamefont {V.}~\bibnamefont {Va{\v{n}}o}},
  \bibinfo {author} {\bibfnamefont {M.}~\bibnamefont {Aapro}}, \bibinfo
  {author} {\bibfnamefont {S.~C.}\ \bibnamefont {Ganguli}}, \bibinfo {author}
  {\bibfnamefont {O.~J.}\ \bibnamefont {Silveira}}, \bibinfo {author}
  {\bibfnamefont {S.}~\bibnamefont {G{\l}odzik}}, \bibinfo {author}
  {\bibfnamefont {A.~S.}\ \bibnamefont {Foster}}, \bibinfo {author}
  {\bibfnamefont {T.}~\bibnamefont {Ojanen}},\ and\ \bibinfo {author}
  {\bibfnamefont {P.}~\bibnamefont {Liljeroth}},\ }\bibfield  {title} {\bibinfo
  {title} {Topological superconductivity in a van der {W}aals
  heterostructure},\ }\href {https://doi.org/10.1038/s41586-020-2989-y}
  {\bibfield  {journal} {\bibinfo  {journal} {Nature}\ }\textbf {\bibinfo
  {volume} {588}},\ \bibinfo {pages} {424} (\bibinfo {year}
  {2020})}\BibitemShut {NoStop}%
\bibitem [{\citenamefont {You}\ \emph {et~al.}(2019)\citenamefont {You},
  \citenamefont {Chen}, \citenamefont {Zhang}, \citenamefont {Sheng},
  \citenamefont {Yang},\ and\ \citenamefont {Su}}]{You2019}%
  \BibitemOpen
  \bibfield  {author} {\bibinfo {author} {\bibfnamefont {J.-Y.}\ \bibnamefont
  {You}}, \bibinfo {author} {\bibfnamefont {C.}~\bibnamefont {Chen}}, \bibinfo
  {author} {\bibfnamefont {Z.}~\bibnamefont {Zhang}}, \bibinfo {author}
  {\bibfnamefont {X.-L.}\ \bibnamefont {Sheng}}, \bibinfo {author}
  {\bibfnamefont {S.~A.}\ \bibnamefont {Yang}},\ and\ \bibinfo {author}
  {\bibfnamefont {G.}~\bibnamefont {Su}},\ }\bibfield  {title} {\bibinfo
  {title} {Two-dimensional {W}eyl half-semimetal and tunable quantum anomalous
  {H}all effect},\ }\href {https://doi.org/10.1103/PhysRevB.100.064408}
  {\bibfield  {journal} {\bibinfo  {journal} {Phys. Rev. B}\ }\textbf {\bibinfo
  {volume} {100}},\ \bibinfo {pages} {064408} (\bibinfo {year}
  {2019})}\BibitemShut {NoStop}%
\bibitem [{\citenamefont {Niu}\ \emph {et~al.}(2019)\citenamefont {Niu},
  \citenamefont {Hanke}, \citenamefont {Buhl}, \citenamefont {Zhang},
  \citenamefont {Plucinski}, \citenamefont {Wortmann}, \citenamefont
  {Bl{\"u}gel}, \citenamefont {Bihlmayer},\ and\ \citenamefont
  {Mokrousov}}]{Niu2019}%
  \BibitemOpen
  \bibfield  {author} {\bibinfo {author} {\bibfnamefont {C.}~\bibnamefont
  {Niu}}, \bibinfo {author} {\bibfnamefont {J.-P.}\ \bibnamefont {Hanke}},
  \bibinfo {author} {\bibfnamefont {P.~M.}\ \bibnamefont {Buhl}}, \bibinfo
  {author} {\bibfnamefont {H.}~\bibnamefont {Zhang}}, \bibinfo {author}
  {\bibfnamefont {L.}~\bibnamefont {Plucinski}}, \bibinfo {author}
  {\bibfnamefont {D.}~\bibnamefont {Wortmann}}, \bibinfo {author}
  {\bibfnamefont {S.}~\bibnamefont {Bl{\"u}gel}}, \bibinfo {author}
  {\bibfnamefont {G.}~\bibnamefont {Bihlmayer}},\ and\ \bibinfo {author}
  {\bibfnamefont {Y.}~\bibnamefont {Mokrousov}},\ }\bibfield  {title} {\bibinfo
  {title} {Mixed topological semimetals driven by orbital complexity in
  two-dimensional ferromagnets},\ }\href
  {https://doi.org/10.1038/s41467-019-10930-6} {\bibfield  {journal} {\bibinfo
  {journal} {Nature Communications}\ }\textbf {\bibinfo {volume} {10}},\
  \bibinfo {pages} {3179} (\bibinfo {year} {2019})}\BibitemShut {NoStop}%
\bibitem [{\citenamefont {Xu}\ \emph {et~al.}(2022)\citenamefont {Xu},
  \citenamefont {Yi}, \citenamefont {Huan}, \citenamefont {Zhao}, \citenamefont
  {Xue},\ and\ \citenamefont {Yang}}]{Xu2022}%
  \BibitemOpen
  \bibfield  {author} {\bibinfo {author} {\bibfnamefont {W.}~\bibnamefont
  {Xu}}, \bibinfo {author} {\bibfnamefont {J.}~\bibnamefont {Yi}}, \bibinfo
  {author} {\bibfnamefont {H.}~\bibnamefont {Huan}}, \bibinfo {author}
  {\bibfnamefont {B.}~\bibnamefont {Zhao}}, \bibinfo {author} {\bibfnamefont
  {Y.}~\bibnamefont {Xue}},\ and\ \bibinfo {author} {\bibfnamefont
  {Z.}~\bibnamefont {Yang}},\ }\bibfield  {title} {\bibinfo {title}
  {Two-dimensional half {C}hern-{W}eyl semimetal with multiple screw axes},\
  }\href {https://doi.org/10.1103/PhysRevB.106.205108} {\bibfield  {journal}
  {\bibinfo  {journal} {Phys. Rev. B}\ }\textbf {\bibinfo {volume} {106}},\
  \bibinfo {pages} {205108} (\bibinfo {year} {2022})}\BibitemShut {NoStop}%
\bibitem [{\citenamefont {Niu}\ \emph {et~al.}(2020)\citenamefont {Niu},
  \citenamefont {Wang}, \citenamefont {Mao}, \citenamefont {Huang},
  \citenamefont {Mokrousov},\ and\ \citenamefont {Dai}}]{Niu2020}%
  \BibitemOpen
  \bibfield  {author} {\bibinfo {author} {\bibfnamefont {C.}~\bibnamefont
  {Niu}}, \bibinfo {author} {\bibfnamefont {H.}~\bibnamefont {Wang}}, \bibinfo
  {author} {\bibfnamefont {N.}~\bibnamefont {Mao}}, \bibinfo {author}
  {\bibfnamefont {B.}~\bibnamefont {Huang}}, \bibinfo {author} {\bibfnamefont
  {Y.}~\bibnamefont {Mokrousov}},\ and\ \bibinfo {author} {\bibfnamefont
  {Y.}~\bibnamefont {Dai}},\ }\bibfield  {title} {\bibinfo {title}
  {Antiferromagnetic topological insulator with nonsymmorphic protection in two
  dimensions},\ }\href {https://doi.org/10.1103/PhysRevLett.124.066401}
  {\bibfield  {journal} {\bibinfo  {journal} {Phys. Rev. Lett.}\ }\textbf
  {\bibinfo {volume} {124}},\ \bibinfo {pages} {066401} (\bibinfo {year}
  {2020})}\BibitemShut {NoStop}%
\bibitem [{\citenamefont {Wang}\ \emph {et~al.}(2020)\citenamefont {Wang},
  \citenamefont {Mao}, \citenamefont {Niu}, \citenamefont {Shen}, \citenamefont
  {Whangbo}, \citenamefont {Huang},\ and\ \citenamefont {Dai}}]{Wang2020}%
  \BibitemOpen
  \bibfield  {author} {\bibinfo {author} {\bibfnamefont {H.}~\bibnamefont
  {Wang}}, \bibinfo {author} {\bibfnamefont {N.}~\bibnamefont {Mao}}, \bibinfo
  {author} {\bibfnamefont {C.}~\bibnamefont {Niu}}, \bibinfo {author}
  {\bibfnamefont {S.}~\bibnamefont {Shen}}, \bibinfo {author} {\bibfnamefont
  {M.-H.}\ \bibnamefont {Whangbo}}, \bibinfo {author} {\bibfnamefont
  {B.}~\bibnamefont {Huang}},\ and\ \bibinfo {author} {\bibfnamefont
  {Y.}~\bibnamefont {Dai}},\ }\bibfield  {title} {\bibinfo {title}
  {Ferromagnetic dual topological insulator in a two-dimensional honeycomb
  lattice},\ }\href {https://doi.org/10.1039/D0MH00803F} {\bibfield  {journal}
  {\bibinfo  {journal} {Mater. Horiz.}\ }\textbf {\bibinfo {volume} {7}},\
  \bibinfo {pages} {2431} (\bibinfo {year} {2020})}\BibitemShut {NoStop}%
\bibitem [{\citenamefont {Mishra}\ and\ \citenamefont
  {Lee}(2018)}]{Mishra2018}%
  \BibitemOpen
  \bibfield  {author} {\bibinfo {author} {\bibfnamefont {A.}~\bibnamefont
  {Mishra}}\ and\ \bibinfo {author} {\bibfnamefont {S.}~\bibnamefont {Lee}},\
  }\bibfield  {title} {\bibinfo {title} {Magnetic {C}hern insulators in a
  monolayer of transition metal trichalcogenides},\ }\href
  {https://doi.org/10.1038/s41598-017-18880-z} {\bibfield  {journal} {\bibinfo
  {journal} {Scientific Reports}\ }\textbf {\bibinfo {volume} {8}},\ \bibinfo
  {pages} {799} (\bibinfo {year} {2018})}\BibitemShut {NoStop}%
\bibitem [{\citenamefont {Sugita}\ \emph {et~al.}(2018)\citenamefont {Sugita},
  \citenamefont {Miyake},\ and\ \citenamefont {Motome}}]{Sugita2018}%
  \BibitemOpen
  \bibfield  {author} {\bibinfo {author} {\bibfnamefont {Y.}~\bibnamefont
  {Sugita}}, \bibinfo {author} {\bibfnamefont {T.}~\bibnamefont {Miyake}},\
  and\ \bibinfo {author} {\bibfnamefont {Y.}~\bibnamefont {Motome}},\
  }\bibfield  {title} {\bibinfo {title} {Multiple {D}irac cones and topological
  magnetism in honeycomb-monolayer transition metal trichalcogenides},\ }\href
  {https://doi.org/10.1103/PhysRevB.97.035125} {\bibfield  {journal} {\bibinfo
  {journal} {Phys. Rev. B}\ }\textbf {\bibinfo {volume} {97}},\ \bibinfo
  {pages} {035125} (\bibinfo {year} {2018})}\BibitemShut {NoStop}%
\bibitem [{\citenamefont {Kang}\ \emph {et~al.}(2023)\citenamefont {Kang},
  \citenamefont {Kang}, \citenamefont {Kim},\ and\ \citenamefont
  {Yu}}]{Kang2023}%
  \BibitemOpen
  \bibfield  {author} {\bibinfo {author} {\bibfnamefont {S.}~\bibnamefont
  {Kang}}, \bibinfo {author} {\bibfnamefont {S.}~\bibnamefont {Kang}}, \bibinfo
  {author} {\bibfnamefont {H.-S.}\ \bibnamefont {Kim}},\ and\ \bibinfo {author}
  {\bibfnamefont {J.}~\bibnamefont {Yu}},\ }\bibfield  {title} {\bibinfo
  {title} {Field-controlled quantum anomalous {H}all effect in electron-doped
  {CrSiTe$_3$} monolayer},\ }\href {https://doi.org/10.1038/s41699-023-00375-3}
  {\bibfield  {journal} {\bibinfo  {journal} {npj 2D Materials and
  Applications}\ }\textbf {\bibinfo {volume} {7}},\ \bibinfo {pages} {13}
  (\bibinfo {year} {2023})}\BibitemShut {NoStop}%
\bibitem [{\citenamefont {Qiao}\ \emph {et~al.}(2010)\citenamefont {Qiao},
  \citenamefont {Yang}, \citenamefont {Feng}, \citenamefont {Tse},
  \citenamefont {Ding}, \citenamefont {Yao}, \citenamefont {Wang},\ and\
  \citenamefont {Niu}}]{Niu2010}%
  \BibitemOpen
  \bibfield  {author} {\bibinfo {author} {\bibfnamefont {Z.}~\bibnamefont
  {Qiao}}, \bibinfo {author} {\bibfnamefont {S.~A.}\ \bibnamefont {Yang}},
  \bibinfo {author} {\bibfnamefont {W.}~\bibnamefont {Feng}}, \bibinfo {author}
  {\bibfnamefont {W.-K.}\ \bibnamefont {Tse}}, \bibinfo {author} {\bibfnamefont
  {J.}~\bibnamefont {Ding}}, \bibinfo {author} {\bibfnamefont {Y.}~\bibnamefont
  {Yao}}, \bibinfo {author} {\bibfnamefont {J.}~\bibnamefont {Wang}},\ and\
  \bibinfo {author} {\bibfnamefont {Q.}~\bibnamefont {Niu}},\ }\bibfield
  {title} {\bibinfo {title} {Quantum anomalous {H}all effect in graphene from
  rashba and exchange effects},\ }\href
  {https://doi.org/10.1103/PhysRevB.82.161414} {\bibfield  {journal} {\bibinfo
  {journal} {Phys. Rev. B}\ }\textbf {\bibinfo {volume} {82}},\ \bibinfo
  {pages} {161414} (\bibinfo {year} {2010})}\BibitemShut {NoStop}%
\bibitem [{\citenamefont {Yang}\ \emph {et~al.}(2011)\citenamefont {Yang},
  \citenamefont {Xu}, \citenamefont {Sheng}, \citenamefont {Wang},
  \citenamefont {Xing},\ and\ \citenamefont {Sheng}}]{Yang2011}%
  \BibitemOpen
  \bibfield  {author} {\bibinfo {author} {\bibfnamefont {Y.}~\bibnamefont
  {Yang}}, \bibinfo {author} {\bibfnamefont {Z.}~\bibnamefont {Xu}}, \bibinfo
  {author} {\bibfnamefont {L.}~\bibnamefont {Sheng}}, \bibinfo {author}
  {\bibfnamefont {B.}~\bibnamefont {Wang}}, \bibinfo {author} {\bibfnamefont
  {D.~Y.}\ \bibnamefont {Xing}},\ and\ \bibinfo {author} {\bibfnamefont
  {D.~N.}\ \bibnamefont {Sheng}},\ }\bibfield  {title} {\bibinfo {title}
  {Time-reversal-symmetry-broken quantum spin {H}all effect},\ }\href
  {https://doi.org/10.1103/PhysRevLett.107.066602} {\bibfield  {journal}
  {\bibinfo  {journal} {Phys. Rev. Lett.}\ }\textbf {\bibinfo {volume} {107}},\
  \bibinfo {pages} {066602} (\bibinfo {year} {2011})}\BibitemShut {NoStop}%
\bibitem [{\citenamefont {Qiao}\ \emph {et~al.}(2011)\citenamefont {Qiao},
  \citenamefont {Tse}, \citenamefont {Jiang}, \citenamefont {Yao},\ and\
  \citenamefont {Niu}}]{Qiao2011}%
  \BibitemOpen
  \bibfield  {author} {\bibinfo {author} {\bibfnamefont {Z.}~\bibnamefont
  {Qiao}}, \bibinfo {author} {\bibfnamefont {W.-K.}\ \bibnamefont {Tse}},
  \bibinfo {author} {\bibfnamefont {H.}~\bibnamefont {Jiang}}, \bibinfo
  {author} {\bibfnamefont {Y.}~\bibnamefont {Yao}},\ and\ \bibinfo {author}
  {\bibfnamefont {Q.}~\bibnamefont {Niu}},\ }\bibfield  {title} {\bibinfo
  {title} {Two-dimensional topological insulator state and topological phase
  transition in bilayer graphene},\ }\href
  {https://doi.org/10.1103/PhysRevLett.107.256801} {\bibfield  {journal}
  {\bibinfo  {journal} {Phys. Rev. Lett.}\ }\textbf {\bibinfo {volume} {107}},\
  \bibinfo {pages} {256801} (\bibinfo {year} {2011})}\BibitemShut {NoStop}%
\bibitem [{\citenamefont {Ezawa}(2012)}]{Ezawa2011}%
  \BibitemOpen
  \bibfield  {author} {\bibinfo {author} {\bibfnamefont {M.}~\bibnamefont
  {Ezawa}},\ }\bibfield  {title} {\bibinfo {title} {Valley-polarized metals and
  quantum anomalous {H}all effect in silicene},\ }\href
  {https://doi.org/10.1103/PhysRevLett.109.055502} {\bibfield  {journal}
  {\bibinfo  {journal} {Phys. Rev. Lett.}\ }\textbf {\bibinfo {volume} {109}},\
  \bibinfo {pages} {055502} (\bibinfo {year} {2012})}\BibitemShut {NoStop}%
\bibitem [{\citenamefont {Ren}\ \emph {et~al.}(2016)\citenamefont {Ren},
  \citenamefont {Zeng}, \citenamefont {Deng}, \citenamefont {Yang},
  \citenamefont {Pan},\ and\ \citenamefont {Qiao}}]{Ren2016}%
  \BibitemOpen
  \bibfield  {author} {\bibinfo {author} {\bibfnamefont {Y.}~\bibnamefont
  {Ren}}, \bibinfo {author} {\bibfnamefont {J.}~\bibnamefont {Zeng}}, \bibinfo
  {author} {\bibfnamefont {X.}~\bibnamefont {Deng}}, \bibinfo {author}
  {\bibfnamefont {F.}~\bibnamefont {Yang}}, \bibinfo {author} {\bibfnamefont
  {H.}~\bibnamefont {Pan}},\ and\ \bibinfo {author} {\bibfnamefont
  {Z.}~\bibnamefont {Qiao}},\ }\bibfield  {title} {\bibinfo {title} {Quantum
  anomalous {H}all effect in atomic crystal layers from in-plane
  magnetization},\ }\href {https://doi.org/10.1103/PhysRevB.94.085411}
  {\bibfield  {journal} {\bibinfo  {journal} {Phys. Rev. B}\ }\textbf {\bibinfo
  {volume} {94}},\ \bibinfo {pages} {085411} (\bibinfo {year}
  {2016})}\BibitemShut {NoStop}%
\bibitem [{\citenamefont {Offidani}\ and\ \citenamefont
  {Ferreira}(2018)}]{Offidani2018}%
  \BibitemOpen
  \bibfield  {author} {\bibinfo {author} {\bibfnamefont {M.}~\bibnamefont
  {Offidani}}\ and\ \bibinfo {author} {\bibfnamefont {A.}~\bibnamefont
  {Ferreira}},\ }\bibfield  {title} {\bibinfo {title} {Anomalous {H}all effect
  in 2d {D}irac materials},\ }\href
  {https://doi.org/10.1103/PhysRevLett.121.126802} {\bibfield  {journal}
  {\bibinfo  {journal} {Phys. Rev. Lett.}\ }\textbf {\bibinfo {volume} {121}},\
  \bibinfo {pages} {126802} (\bibinfo {year} {2018})}\BibitemShut {NoStop}%
\bibitem [{\citenamefont {H\"ogl}\ \emph {et~al.}(2020)\citenamefont {H\"ogl},
  \citenamefont {Frank}, \citenamefont {Zollner}, \citenamefont {Kochan},
  \citenamefont {Gmitra},\ and\ \citenamefont {Fabian}}]{Hogl2018}%
  \BibitemOpen
  \bibfield  {author} {\bibinfo {author} {\bibfnamefont {P.}~\bibnamefont
  {H\"ogl}}, \bibinfo {author} {\bibfnamefont {T.}~\bibnamefont {Frank}},
  \bibinfo {author} {\bibfnamefont {K.}~\bibnamefont {Zollner}}, \bibinfo
  {author} {\bibfnamefont {D.}~\bibnamefont {Kochan}}, \bibinfo {author}
  {\bibfnamefont {M.}~\bibnamefont {Gmitra}},\ and\ \bibinfo {author}
  {\bibfnamefont {J.}~\bibnamefont {Fabian}},\ }\bibfield  {title} {\bibinfo
  {title} {Quantum anomalous {H}all effects in graphene from proximity-induced
  uniform and staggered spin-orbit and exchange coupling},\ }\href
  {https://doi.org/10.1103/PhysRevLett.124.136403} {\bibfield  {journal}
  {\bibinfo  {journal} {Phys. Rev. Lett.}\ }\textbf {\bibinfo {volume} {124}},\
  \bibinfo {pages} {136403} (\bibinfo {year} {2020})}\BibitemShut {NoStop}%
\bibitem [{\citenamefont {Zou}\ \emph {et~al.}(2020)\citenamefont {Zou},
  \citenamefont {Zhan}, \citenamefont {Zheng}, \citenamefont {Wu},
  \citenamefont {Fan},\ and\ \citenamefont {Wang}}]{Zou2020}%
  \BibitemOpen
  \bibfield  {author} {\bibinfo {author} {\bibfnamefont {R.}~\bibnamefont
  {Zou}}, \bibinfo {author} {\bibfnamefont {F.}~\bibnamefont {Zhan}}, \bibinfo
  {author} {\bibfnamefont {B.}~\bibnamefont {Zheng}}, \bibinfo {author}
  {\bibfnamefont {X.}~\bibnamefont {Wu}}, \bibinfo {author} {\bibfnamefont
  {J.}~\bibnamefont {Fan}},\ and\ \bibinfo {author} {\bibfnamefont
  {R.}~\bibnamefont {Wang}},\ }\bibfield  {title} {\bibinfo {title} {Intrinsic
  quantum anomalous {H}all phase induced by proximity in the van der {W}aals
  heterostructure germanene/{Cr$_2$Ge$_2$Te$_6$}},\ }\href
  {https://doi.org/10.1103/PhysRevB.101.161108} {\bibfield  {journal} {\bibinfo
   {journal} {Phys. Rev. B}\ }\textbf {\bibinfo {volume} {101}},\ \bibinfo
  {pages} {161108} (\bibinfo {year} {2020})}\BibitemShut {NoStop}%
\bibitem [{\citenamefont {Vila}\ \emph {et~al.}(2021)\citenamefont {Vila},
  \citenamefont {Garcia},\ and\ \citenamefont {Roche}}]{Vila2021}%
  \BibitemOpen
  \bibfield  {author} {\bibinfo {author} {\bibfnamefont {M.}~\bibnamefont
  {Vila}}, \bibinfo {author} {\bibfnamefont {J.~H.}\ \bibnamefont {Garcia}},\
  and\ \bibinfo {author} {\bibfnamefont {S.}~\bibnamefont {Roche}},\ }\bibfield
   {title} {\bibinfo {title} {Valley-polarized quantum anomalous {H}all phase
  in bilayer graphene with layer-dependent proximity effects},\ }\href
  {https://doi.org/10.1103/PhysRevB.104.L161113} {\bibfield  {journal}
  {\bibinfo  {journal} {Phys. Rev. B}\ }\textbf {\bibinfo {volume} {104}},\
  \bibinfo {pages} {L161113} (\bibinfo {year} {2021})}\BibitemShut {NoStop}%
\bibitem [{\citenamefont {Sheremetyeva}\ \emph {et~al.}(2023)\citenamefont
  {Sheremetyeva}, \citenamefont {Na}, \citenamefont {Saraf}, \citenamefont
  {Griffin},\ and\ \citenamefont {Hautier}}]{Natalya2023}%
  \BibitemOpen
  \bibfield  {author} {\bibinfo {author} {\bibfnamefont {N.}~\bibnamefont
  {Sheremetyeva}}, \bibinfo {author} {\bibfnamefont {I.}~\bibnamefont {Na}},
  \bibinfo {author} {\bibfnamefont {A.}~\bibnamefont {Saraf}}, \bibinfo
  {author} {\bibfnamefont {S.~M.}\ \bibnamefont {Griffin}},\ and\ \bibinfo
  {author} {\bibfnamefont {G.}~\bibnamefont {Hautier}},\ }\bibfield  {title}
  {\bibinfo {title} {Prediction of topological phases in metastable
  ferromagnetic {M}$\mathrm{P}{X}_{3}$ monolayers},\ }\href
  {https://doi.org/10.1103/PhysRevB.107.115104} {\bibfield  {journal} {\bibinfo
   {journal} {Phys. Rev. B}\ }\textbf {\bibinfo {volume} {107}},\ \bibinfo
  {pages} {115104} (\bibinfo {year} {2023})}\BibitemShut {NoStop}%
\bibitem [{\citenamefont {Shiozaki}\ and\ \citenamefont
  {Sato}(2014)}]{shiozaki2014}%
  \BibitemOpen
  \bibfield  {author} {\bibinfo {author} {\bibfnamefont {K.}~\bibnamefont
  {Shiozaki}}\ and\ \bibinfo {author} {\bibfnamefont {M.}~\bibnamefont
  {Sato}},\ }\bibfield  {title} {\bibinfo {title} {Topology of crystalline
  insulators and superconductors},\ }\href
  {https://doi.org/10.1103/PhysRevB.90.165114} {\bibfield  {journal} {\bibinfo
  {journal} {Phys. Rev. B}\ }\textbf {\bibinfo {volume} {90}},\ \bibinfo
  {pages} {165114} (\bibinfo {year} {2014})}\BibitemShut {NoStop}%
\bibitem [{\citenamefont {Wang}\ \emph {et~al.}(2018)\citenamefont {Wang},
  \citenamefont {Shifa}, \citenamefont {Yu}, \citenamefont {He}, \citenamefont
  {Liu}, \citenamefont {Wang}, \citenamefont {Wang}, \citenamefont {Zhan},
  \citenamefont {Lou}, \citenamefont {Xia},\ and\ \citenamefont
  {He}}]{Wang2018}%
  \BibitemOpen
  \bibfield  {author} {\bibinfo {author} {\bibfnamefont {F.}~\bibnamefont
  {Wang}}, \bibinfo {author} {\bibfnamefont {T.~A.}\ \bibnamefont {Shifa}},
  \bibinfo {author} {\bibfnamefont {P.}~\bibnamefont {Yu}}, \bibinfo {author}
  {\bibfnamefont {P.}~\bibnamefont {He}}, \bibinfo {author} {\bibfnamefont
  {Y.}~\bibnamefont {Liu}}, \bibinfo {author} {\bibfnamefont {F.}~\bibnamefont
  {Wang}}, \bibinfo {author} {\bibfnamefont {Z.}~\bibnamefont {Wang}}, \bibinfo
  {author} {\bibfnamefont {X.}~\bibnamefont {Zhan}}, \bibinfo {author}
  {\bibfnamefont {X.}~\bibnamefont {Lou}}, \bibinfo {author} {\bibfnamefont
  {F.}~\bibnamefont {Xia}},\ and\ \bibinfo {author} {\bibfnamefont
  {J.}~\bibnamefont {He}},\ }\bibfield  {title} {\bibinfo {title} {New
  frontiers on van der waals layered metal phosphorous trichalcogenides},\
  }\href {https://doi.org/10.1002/adfm.201802151} {\bibfield  {journal}
  {\bibinfo  {journal} {Advanced Functional Materials}\ }\textbf {\bibinfo
  {volume} {28}},\ \bibinfo {pages} {1802151} (\bibinfo {year}
  {2018})}\BibitemShut {NoStop}%
\bibitem [{\citenamefont {Sivadas}\ \emph {et~al.}(2015)\citenamefont
  {Sivadas}, \citenamefont {Daniels}, \citenamefont {Swendsen}, \citenamefont
  {Okamoto},\ and\ \citenamefont {Xiao}}]{Sivadas2015}%
  \BibitemOpen
  \bibfield  {author} {\bibinfo {author} {\bibfnamefont {N.}~\bibnamefont
  {Sivadas}}, \bibinfo {author} {\bibfnamefont {M.~W.}\ \bibnamefont
  {Daniels}}, \bibinfo {author} {\bibfnamefont {R.~H.}\ \bibnamefont
  {Swendsen}}, \bibinfo {author} {\bibfnamefont {S.}~\bibnamefont {Okamoto}},\
  and\ \bibinfo {author} {\bibfnamefont {D.}~\bibnamefont {Xiao}},\ }\bibfield
  {title} {\bibinfo {title} {Magnetic ground state of semiconducting
  transition-metal trichalcogenide monolayers},\ }\href
  {https://doi.org/10.1103/PhysRevB.91.235425} {\bibfield  {journal} {\bibinfo
  {journal} {Phys. Rev. B}\ }\textbf {\bibinfo {volume} {91}},\ \bibinfo
  {pages} {235425} (\bibinfo {year} {2015})}\BibitemShut {NoStop}%
\bibitem [{\citenamefont {Mak}\ \emph {et~al.}(2019)\citenamefont {Mak},
  \citenamefont {Shan},\ and\ \citenamefont {Ralph}}]{Mak2019}%
  \BibitemOpen
  \bibfield  {author} {\bibinfo {author} {\bibfnamefont {K.~F.}\ \bibnamefont
  {Mak}}, \bibinfo {author} {\bibfnamefont {J.}~\bibnamefont {Shan}},\ and\
  \bibinfo {author} {\bibfnamefont {D.~C.}\ \bibnamefont {Ralph}},\ }\bibfield
  {title} {\bibinfo {title} {Probing and controlling magnetic states in 2d
  layered magnetic materials},\ }\href
  {https://doi.org/10.1038/s42254-019-0110-y} {\bibfield  {journal} {\bibinfo
  {journal} {Nature Reviews Physics}\ }\textbf {\bibinfo {volume} {1}},\
  \bibinfo {pages} {646} (\bibinfo {year} {2019})}\BibitemShut {NoStop}%
\bibitem [{\citenamefont {Zhang}\ \emph {et~al.}(2016)\citenamefont {Zhang},
  \citenamefont {Zhao}, \citenamefont {Wu}, \citenamefont {Jing},\ and\
  \citenamefont {Zhou}}]{Zhang2016}%
  \BibitemOpen
  \bibfield  {author} {\bibinfo {author} {\bibfnamefont {X.}~\bibnamefont
  {Zhang}}, \bibinfo {author} {\bibfnamefont {X.}~\bibnamefont {Zhao}},
  \bibinfo {author} {\bibfnamefont {D.}~\bibnamefont {Wu}}, \bibinfo {author}
  {\bibfnamefont {Y.}~\bibnamefont {Jing}},\ and\ \bibinfo {author}
  {\bibfnamefont {Z.}~\bibnamefont {Zhou}},\ }\bibfield  {title} {\bibinfo
  {title} {{MnPSe$_3$} monolayer: A promising 2{D} visible-light
  photohydrolytic catalyst with high carrier mobility},\ }\href
  {https://doi.org/https://doi.org/10.1002/advs.201600062} {\bibfield
  {journal} {\bibinfo  {journal} {Advanced Science}\ }\textbf {\bibinfo
  {volume} {3}},\ \bibinfo {pages} {1600062} (\bibinfo {year}
  {2016})}\BibitemShut {NoStop}%
\bibitem [{\citenamefont {Chittari}\ \emph {et~al.}(2016)\citenamefont
  {Chittari}, \citenamefont {Park}, \citenamefont {Lee}, \citenamefont {Han},
  \citenamefont {MacDonald}, \citenamefont {Hwang},\ and\ \citenamefont
  {Jung}}]{Macdonald2016}%
  \BibitemOpen
  \bibfield  {author} {\bibinfo {author} {\bibfnamefont {B.~L.}\ \bibnamefont
  {Chittari}}, \bibinfo {author} {\bibfnamefont {Y.}~\bibnamefont {Park}},
  \bibinfo {author} {\bibfnamefont {D.}~\bibnamefont {Lee}}, \bibinfo {author}
  {\bibfnamefont {M.}~\bibnamefont {Han}}, \bibinfo {author} {\bibfnamefont
  {A.~H.}\ \bibnamefont {MacDonald}}, \bibinfo {author} {\bibfnamefont
  {E.}~\bibnamefont {Hwang}},\ and\ \bibinfo {author} {\bibfnamefont
  {J.}~\bibnamefont {Jung}},\ }\bibfield  {title} {\bibinfo {title} {Electronic
  and magnetic properties of single-layer {M}$\mathrm{P}{X}_{3}$ metal
  phosphorous trichalcogenides},\ }\href
  {https://doi.org/10.1103/PhysRevB.94.184428} {\bibfield  {journal} {\bibinfo
  {journal} {Phys. Rev. B}\ }\textbf {\bibinfo {volume} {94}},\ \bibinfo
  {pages} {184428} (\bibinfo {year} {2016})}\BibitemShut {NoStop}%
\bibitem [{\citenamefont {Li}\ \emph {et~al.}(2014)\citenamefont {Li},
  \citenamefont {Wu},\ and\ \citenamefont {Yang}}]{Li2014}%
  \BibitemOpen
  \bibfield  {author} {\bibinfo {author} {\bibfnamefont {X.}~\bibnamefont
  {Li}}, \bibinfo {author} {\bibfnamefont {X.}~\bibnamefont {Wu}},\ and\
  \bibinfo {author} {\bibfnamefont {J.}~\bibnamefont {Yang}},\ }\bibfield
  {title} {\bibinfo {title} {Half-metallicity in {MnPSe$_3$} exfoliated
  nanosheet with carrier doping},\ }\href {https://doi.org/10.1021/ja505097m}
  {\bibfield  {journal} {\bibinfo  {journal} {Journal of the American Chemical
  Society}\ }\textbf {\bibinfo {volume} {136}},\ \bibinfo {pages} {11065}
  (\bibinfo {year} {2014})},\ \bibinfo {note} {pMID: 25036853},\ \Eprint
  {https://arxiv.org/abs/https://doi.org/10.1021/ja505097m}
  {https://doi.org/10.1021/ja505097m} \BibitemShut {NoStop}%
\bibitem [{\citenamefont {Pei}\ \emph {et~al.}(2018)\citenamefont {Pei},
  \citenamefont {Wang}, \citenamefont {Zou},\ and\ \citenamefont
  {Mi}}]{Pei2018}%
  \BibitemOpen
  \bibfield  {author} {\bibinfo {author} {\bibfnamefont {Q.}~\bibnamefont
  {Pei}}, \bibinfo {author} {\bibfnamefont {X.-C.}\ \bibnamefont {Wang}},
  \bibinfo {author} {\bibfnamefont {J.-J.}\ \bibnamefont {Zou}},\ and\ \bibinfo
  {author} {\bibfnamefont {W.-B.}\ \bibnamefont {Mi}},\ }\bibfield  {title}
  {\bibinfo {title} {Tunable electronic structure and magnetic coupling in
  strained two-dimensional semiconductor {M}n{P}{S}e$_3$},\ }\href
  {https://doi.org/10.1007/s11467-018-0796-9} {\bibfield  {journal} {\bibinfo
  {journal} {Frontiers of Physics}\ }\textbf {\bibinfo {volume} {13}},\
  \bibinfo {pages} {137105} (\bibinfo {year} {2018})}\BibitemShut {NoStop}%
\bibitem [{\citenamefont {Kohn}\ and\ \citenamefont
  {Sham}(1965)}]{KohnSham1965}%
  \BibitemOpen
  \bibfield  {author} {\bibinfo {author} {\bibfnamefont {W.}~\bibnamefont
  {Kohn}}\ and\ \bibinfo {author} {\bibfnamefont {L.~J.}\ \bibnamefont
  {Sham}},\ }\bibfield  {title} {\bibinfo {title} {Self-consistent equations
  including exchange and correlation effects},\ }\href
  {https://doi.org/10.1103/PhysRev.140.A1133} {\bibfield  {journal} {\bibinfo
  {journal} {Phys. Rev.}\ }\textbf {\bibinfo {volume} {140}},\ \bibinfo {pages}
  {A1133} (\bibinfo {year} {1965})}\BibitemShut {NoStop}%
\bibitem [{\citenamefont {Kresse}\ and\ \citenamefont
  {Furthm\"uller}(1996)}]{Kresse1996}%
  \BibitemOpen
  \bibfield  {author} {\bibinfo {author} {\bibfnamefont {G.}~\bibnamefont
  {Kresse}}\ and\ \bibinfo {author} {\bibfnamefont {J.}~\bibnamefont
  {Furthm\"uller}},\ }\bibfield  {title} {\bibinfo {title} {Efficient iterative
  schemes for ab initio total-energy calculations using a plane-wave basis
  set},\ }\href {https://doi.org/10.1103/PhysRevB.54.11169} {\bibfield
  {journal} {\bibinfo  {journal} {Phys. Rev. B}\ }\textbf {\bibinfo {volume}
  {54}},\ \bibinfo {pages} {11169} (\bibinfo {year} {1996})}\BibitemShut
  {NoStop}%
\bibitem [{\citenamefont {Kresse}\ and\ \citenamefont
  {Joubert}(1999)}]{Kresse1999}%
  \BibitemOpen
  \bibfield  {author} {\bibinfo {author} {\bibfnamefont {G.}~\bibnamefont
  {Kresse}}\ and\ \bibinfo {author} {\bibfnamefont {D.}~\bibnamefont
  {Joubert}},\ }\bibfield  {title} {\bibinfo {title} {From ultrasoft
  pseudopotentials to the projector augmented-wave method},\ }\href
  {https://doi.org/10.1103/PhysRevB.59.1758} {\bibfield  {journal} {\bibinfo
  {journal} {Phys. Rev. B}\ }\textbf {\bibinfo {volume} {59}},\ \bibinfo
  {pages} {1758} (\bibinfo {year} {1999})}\BibitemShut {NoStop}%
\bibitem [{\citenamefont {Perdew}\ \emph {et~al.}(1996)\citenamefont {Perdew},
  \citenamefont {Burke},\ and\ \citenamefont {Ernzerhof}}]{Perdew1997}%
  \BibitemOpen
  \bibfield  {author} {\bibinfo {author} {\bibfnamefont {J.~P.}\ \bibnamefont
  {Perdew}}, \bibinfo {author} {\bibfnamefont {K.}~\bibnamefont {Burke}},\ and\
  \bibinfo {author} {\bibfnamefont {M.}~\bibnamefont {Ernzerhof}},\ }\bibfield
  {title} {\bibinfo {title} {Generalized gradient approximation made simple},\
  }\href {https://doi.org/10.1103/PhysRevLett.77.3865} {\bibfield  {journal}
  {\bibinfo  {journal} {Phys. Rev. Lett.}\ }\textbf {\bibinfo {volume} {77}},\
  \bibinfo {pages} {3865} (\bibinfo {year} {1996})}\BibitemShut {NoStop}%
\bibitem [{\citenamefont {Bl\"ochl}(1994)}]{Blochl1994}%
  \BibitemOpen
  \bibfield  {author} {\bibinfo {author} {\bibfnamefont {P.~E.}\ \bibnamefont
  {Bl\"ochl}},\ }\bibfield  {title} {\bibinfo {title} {Projector augmented-wave
  method},\ }\href {https://doi.org/10.1103/PhysRevB.50.17953} {\bibfield
  {journal} {\bibinfo  {journal} {Phys. Rev. B}\ }\textbf {\bibinfo {volume}
  {50}},\ \bibinfo {pages} {17953} (\bibinfo {year} {1994})}\BibitemShut
  {NoStop}%
\bibitem [{\citenamefont {Souza}\ \emph {et~al.}(2001)\citenamefont {Souza},
  \citenamefont {Marzari},\ and\ \citenamefont {Vanderbilt}}]{Souza2001}%
  \BibitemOpen
  \bibfield  {author} {\bibinfo {author} {\bibfnamefont {I.}~\bibnamefont
  {Souza}}, \bibinfo {author} {\bibfnamefont {N.}~\bibnamefont {Marzari}},\
  and\ \bibinfo {author} {\bibfnamefont {D.}~\bibnamefont {Vanderbilt}},\
  }\bibfield  {title} {\bibinfo {title} {Maximally localized {W}annier
  functions for entangled energy bands},\ }\href
  {https://doi.org/10.1103/PhysRevB.65.035109} {\bibfield  {journal} {\bibinfo
  {journal} {Phys. Rev. B}\ }\textbf {\bibinfo {volume} {65}},\ \bibinfo
  {pages} {035109} (\bibinfo {year} {2001})}\BibitemShut {NoStop}%
\bibitem [{\citenamefont {Wang}\ \emph {et~al.}(2006)\citenamefont {Wang},
  \citenamefont {Yates}, \citenamefont {Souza},\ and\ \citenamefont
  {Vanderbilt}}]{Wang2007}%
  \BibitemOpen
  \bibfield  {author} {\bibinfo {author} {\bibfnamefont {X.}~\bibnamefont
  {Wang}}, \bibinfo {author} {\bibfnamefont {J.~R.}\ \bibnamefont {Yates}},
  \bibinfo {author} {\bibfnamefont {I.}~\bibnamefont {Souza}},\ and\ \bibinfo
  {author} {\bibfnamefont {D.}~\bibnamefont {Vanderbilt}},\ }\bibfield  {title}
  {\bibinfo {title} {Ab initio calculation of the anomalous {H}all conductivity
  by {W}annier interpolation},\ }\href
  {https://doi.org/10.1103/PhysRevB.74.195118} {\bibfield  {journal} {\bibinfo
  {journal} {Phys. Rev. B}\ }\textbf {\bibinfo {volume} {74}},\ \bibinfo
  {pages} {195118} (\bibinfo {year} {2006})}\BibitemShut {NoStop}%
\bibitem [{\citenamefont {Marzari}\ \emph {et~al.}(2012)\citenamefont
  {Marzari}, \citenamefont {Mostofi}, \citenamefont {Yates}, \citenamefont
  {Souza},\ and\ \citenamefont {Vanderbilt}}]{Marzari2012}%
  \BibitemOpen
  \bibfield  {author} {\bibinfo {author} {\bibfnamefont {N.}~\bibnamefont
  {Marzari}}, \bibinfo {author} {\bibfnamefont {A.~A.}\ \bibnamefont
  {Mostofi}}, \bibinfo {author} {\bibfnamefont {J.~R.}\ \bibnamefont {Yates}},
  \bibinfo {author} {\bibfnamefont {I.}~\bibnamefont {Souza}},\ and\ \bibinfo
  {author} {\bibfnamefont {D.}~\bibnamefont {Vanderbilt}},\ }\bibfield  {title}
  {\bibinfo {title} {Maximally localized {W}annier functions: Theory and
  applications},\ }\href {https://doi.org/10.1103/RevModPhys.84.1419}
  {\bibfield  {journal} {\bibinfo  {journal} {Rev. Mod. Phys.}\ }\textbf
  {\bibinfo {volume} {84}},\ \bibinfo {pages} {1419} (\bibinfo {year}
  {2012})}\BibitemShut {NoStop}%
\bibitem [{\citenamefont {Kane}\ and\ \citenamefont {Mele}(2005)}]{Kane2005}%
  \BibitemOpen
  \bibfield  {author} {\bibinfo {author} {\bibfnamefont {C.~L.}\ \bibnamefont
  {Kane}}\ and\ \bibinfo {author} {\bibfnamefont {E.~J.}\ \bibnamefont
  {Mele}},\ }\bibfield  {title} {\bibinfo {title} {Quantum spin {H}all effect
  in graphene},\ }\href {https://doi.org/10.1103/PhysRevLett.95.226801}
  {\bibfield  {journal} {\bibinfo  {journal} {Phys. Rev. Lett.}\ }\textbf
  {\bibinfo {volume} {95}},\ \bibinfo {pages} {226801} (\bibinfo {year}
  {2005})}\BibitemShut {NoStop}%
\bibitem [{\citenamefont {Gmitra}\ \emph {et~al.}(2009)\citenamefont {Gmitra},
  \citenamefont {Konschuh}, \citenamefont {Ertler}, \citenamefont
  {Ambrosch-Draxl},\ and\ \citenamefont {Fabian}}]{Gmitra2009}%
  \BibitemOpen
  \bibfield  {author} {\bibinfo {author} {\bibfnamefont {M.}~\bibnamefont
  {Gmitra}}, \bibinfo {author} {\bibfnamefont {S.}~\bibnamefont {Konschuh}},
  \bibinfo {author} {\bibfnamefont {C.}~\bibnamefont {Ertler}}, \bibinfo
  {author} {\bibfnamefont {C.}~\bibnamefont {Ambrosch-Draxl}},\ and\ \bibinfo
  {author} {\bibfnamefont {J.}~\bibnamefont {Fabian}},\ }\bibfield  {title}
  {\bibinfo {title} {Band-structure topologies of graphene: Spin-orbit coupling
  effects from first principles},\ }\href
  {https://doi.org/10.1103/PhysRevB.80.235431} {\bibfield  {journal} {\bibinfo
  {journal} {Phys. Rev. B}\ }\textbf {\bibinfo {volume} {80}},\ \bibinfo
  {pages} {235431} (\bibinfo {year} {2009})}\BibitemShut {NoStop}%
\bibitem [{\citenamefont {Konschuh}\ \emph {et~al.}(2010)\citenamefont
  {Konschuh}, \citenamefont {Gmitra},\ and\ \citenamefont
  {Fabian}}]{Konschuh2010}%
  \BibitemOpen
  \bibfield  {author} {\bibinfo {author} {\bibfnamefont {S.}~\bibnamefont
  {Konschuh}}, \bibinfo {author} {\bibfnamefont {M.}~\bibnamefont {Gmitra}},\
  and\ \bibinfo {author} {\bibfnamefont {J.}~\bibnamefont {Fabian}},\
  }\bibfield  {title} {\bibinfo {title} {Tight-binding theory of the spin-orbit
  coupling in graphene},\ }\href {https://doi.org/10.1103/PhysRevB.82.245412}
  {\bibfield  {journal} {\bibinfo  {journal} {Phys. Rev. B}\ }\textbf {\bibinfo
  {volume} {82}},\ \bibinfo {pages} {245412} (\bibinfo {year}
  {2010})}\BibitemShut {NoStop}%
\bibitem [{\citenamefont {Aroyo}\ \emph {et~al.}(2011)\citenamefont {Aroyo},
  \citenamefont {Perez-Mato}, \citenamefont {Orobengoa}, \citenamefont {Tasci},
  \citenamefont {De~La~Flor},\ and\ \citenamefont {Kirov}}]{Aroyo2011}%
  \BibitemOpen
  \bibfield  {author} {\bibinfo {author} {\bibfnamefont {M.}~\bibnamefont
  {Aroyo}}, \bibinfo {author} {\bibfnamefont {J.}~\bibnamefont {Perez-Mato}},
  \bibinfo {author} {\bibfnamefont {D.}~\bibnamefont {Orobengoa}}, \bibinfo
  {author} {\bibfnamefont {E.}~\bibnamefont {Tasci}}, \bibinfo {author}
  {\bibfnamefont {G.}~\bibnamefont {De~La~Flor}},\ and\ \bibinfo {author}
  {\bibfnamefont {A.}~\bibnamefont {Kirov}},\ }\bibfield  {title} {\bibinfo
  {title} {Crystallography online: Bilbao crystallographic server},\ }\href
  {https://www.scopus.com/inward/record.uri?eid=2-s2.0-80955140447&partnerID=40&md5=488772b9e21d2636a3952f66ae80ae84}
  {\bibfield  {journal} {\bibinfo  {journal} {Bulgarian Chemical
  Communications}\ }\textbf {\bibinfo {volume} {43}},\ \bibinfo {pages} {183
  – 197} (\bibinfo {year} {2011})}\BibitemShut {NoStop}%
\bibitem [{\citenamefont {Chiu}\ \emph {et~al.}(2016)\citenamefont {Chiu},
  \citenamefont {Teo}, \citenamefont {Schnyder},\ and\ \citenamefont
  {Ryu}}]{Chiu2016}%
  \BibitemOpen
  \bibfield  {author} {\bibinfo {author} {\bibfnamefont {C.-K.}\ \bibnamefont
  {Chiu}}, \bibinfo {author} {\bibfnamefont {J.~C.~Y.}\ \bibnamefont {Teo}},
  \bibinfo {author} {\bibfnamefont {A.~P.}\ \bibnamefont {Schnyder}},\ and\
  \bibinfo {author} {\bibfnamefont {S.}~\bibnamefont {Ryu}},\ }\bibfield
  {title} {\bibinfo {title} {Classification of topological quantum matter with
  symmetries},\ }\href {https://doi.org/10.1103/RevModPhys.88.035005}
  {\bibfield  {journal} {\bibinfo  {journal} {Rev. Mod. Phys.}\ }\textbf
  {\bibinfo {volume} {88}},\ \bibinfo {pages} {035005} (\bibinfo {year}
  {2016})}\BibitemShut {NoStop}%
\bibitem [{\citenamefont {Schnyder}(2020)}]{Andreas2020}%
  \BibitemOpen
  \bibfield  {author} {\bibinfo {author} {\bibfnamefont {A.}~\bibnamefont
  {Schnyder}},\ }\bibinfo {title} {Topological semimetals},\ in\ \href@noop {}
  {\emph {\bibinfo {booktitle} {Topology, entanglement, and strong
  correlations}}},\ \bibinfo {editor} {edited by\ \bibinfo {editor}
  {\bibfnamefont {E.}~\bibnamefont {Pavarini}}\ and\ \bibinfo {editor}
  {\bibfnamefont {E.}~\bibnamefont {Koch}}}\ (\bibinfo  {publisher}
  {Forschungszentrum Jülich GmbH},\ \bibinfo {address} {J\"{u}lich},\ \bibinfo
  {year} {2020})\ Chap.~\bibinfo {chapter} {11}\BibitemShut {NoStop}%
\bibitem [{\citenamefont {Young}\ and\ \citenamefont {Kane}(2015)}]{Young2015}%
  \BibitemOpen
  \bibfield  {author} {\bibinfo {author} {\bibfnamefont {S.~M.}\ \bibnamefont
  {Young}}\ and\ \bibinfo {author} {\bibfnamefont {C.~L.}\ \bibnamefont
  {Kane}},\ }\bibfield  {title} {\bibinfo {title} {{D}irac semimetals in two
  dimensions},\ }\href {https://doi.org/10.1103/PhysRevLett.115.126803}
  {\bibfield  {journal} {\bibinfo  {journal} {Phys. Rev. Lett.}\ }\textbf
  {\bibinfo {volume} {115}},\ \bibinfo {pages} {126803} (\bibinfo {year}
  {2015})}\BibitemShut {NoStop}%
\bibitem [{\citenamefont {Varjas}\ \emph {et~al.}(2018)\citenamefont {Varjas},
  \citenamefont {Rosdahl},\ and\ \citenamefont {Akhmerov}}]{Varjas2018}%
  \BibitemOpen
  \bibfield  {author} {\bibinfo {author} {\bibfnamefont {D.}~\bibnamefont
  {Varjas}}, \bibinfo {author} {\bibfnamefont {T.~{\"O}.}\ \bibnamefont
  {Rosdahl}},\ and\ \bibinfo {author} {\bibfnamefont {A.~R.}\ \bibnamefont
  {Akhmerov}},\ }\bibfield  {title} {\bibinfo {title} {Qsymm: algorithmic
  symmetry finding and symmetric hamiltonian generation},\ }\href
  {https://doi.org/10.1088/1367-2630/aadf67} {\bibfield  {journal} {\bibinfo
  {journal} {New Journal of Physics}\ }\textbf {\bibinfo {volume} {20}},\
  \bibinfo {pages} {093026} (\bibinfo {year} {2018})}\BibitemShut {NoStop}%
\bibitem [{\citenamefont {Benalcazar}\ \emph {et~al.}(2019)\citenamefont
  {Benalcazar}, \citenamefont {Li},\ and\ \citenamefont
  {Hughes}}]{Benalcazar2018}%
  \BibitemOpen
  \bibfield  {author} {\bibinfo {author} {\bibfnamefont {W.~A.}\ \bibnamefont
  {Benalcazar}}, \bibinfo {author} {\bibfnamefont {T.}~\bibnamefont {Li}},\
  and\ \bibinfo {author} {\bibfnamefont {T.~L.}\ \bibnamefont {Hughes}},\
  }\bibfield  {title} {\bibinfo {title} {Quantization of fractional corner
  charge in ${C}_{n}$-symmetric higher-order topological crystalline
  insulators},\ }\href {https://doi.org/10.1103/PhysRevB.99.245151} {\bibfield
  {journal} {\bibinfo  {journal} {Phys. Rev. B}\ }\textbf {\bibinfo {volume}
  {99}},\ \bibinfo {pages} {245151} (\bibinfo {year} {2019})}\BibitemShut
  {NoStop}%
\bibitem [{\citenamefont {Thouless}\ \emph {et~al.}(1982)\citenamefont
  {Thouless}, \citenamefont {Kohmoto}, \citenamefont {Nightingale},\ and\
  \citenamefont {den Nijs}}]{Thouless1982}%
  \BibitemOpen
  \bibfield  {author} {\bibinfo {author} {\bibfnamefont {D.~J.}\ \bibnamefont
  {Thouless}}, \bibinfo {author} {\bibfnamefont {M.}~\bibnamefont {Kohmoto}},
  \bibinfo {author} {\bibfnamefont {M.~P.}\ \bibnamefont {Nightingale}},\ and\
  \bibinfo {author} {\bibfnamefont {M.}~\bibnamefont {den Nijs}},\ }\bibfield
  {title} {\bibinfo {title} {Quantized {H}all conductance in a two-dimensional
  periodic potential},\ }\href {https://doi.org/10.1103/PhysRevLett.49.405}
  {\bibfield  {journal} {\bibinfo  {journal} {Phys. Rev. Lett.}\ }\textbf
  {\bibinfo {volume} {49}},\ \bibinfo {pages} {405} (\bibinfo {year}
  {1982})}\BibitemShut {NoStop}%
\bibitem [{\citenamefont {Xiao}\ \emph {et~al.}(2010)\citenamefont {Xiao},
  \citenamefont {Chang},\ and\ \citenamefont {Niu}}]{Dixiao2010}%
  \BibitemOpen
  \bibfield  {author} {\bibinfo {author} {\bibfnamefont {D.}~\bibnamefont
  {Xiao}}, \bibinfo {author} {\bibfnamefont {M.-C.}\ \bibnamefont {Chang}},\
  and\ \bibinfo {author} {\bibfnamefont {Q.}~\bibnamefont {Niu}},\ }\bibfield
  {title} {\bibinfo {title} {Berry phase effects on electronic properties},\
  }\href {https://doi.org/10.1103/RevModPhys.82.1959} {\bibfield  {journal}
  {\bibinfo  {journal} {Rev. Mod. Phys.}\ }\textbf {\bibinfo {volume} {82}},\
  \bibinfo {pages} {1959} (\bibinfo {year} {2010})}\BibitemShut {NoStop}%
\bibitem [{\citenamefont {Kruthoff}\ \emph {et~al.}(2017)\citenamefont
  {Kruthoff}, \citenamefont {de~Boer}, \citenamefont {van Wezel}, \citenamefont
  {Kane},\ and\ \citenamefont {Slager}}]{Kruthoff2017}%
  \BibitemOpen
  \bibfield  {author} {\bibinfo {author} {\bibfnamefont {J.}~\bibnamefont
  {Kruthoff}}, \bibinfo {author} {\bibfnamefont {J.}~\bibnamefont {de~Boer}},
  \bibinfo {author} {\bibfnamefont {J.}~\bibnamefont {van Wezel}}, \bibinfo
  {author} {\bibfnamefont {C.~L.}\ \bibnamefont {Kane}},\ and\ \bibinfo
  {author} {\bibfnamefont {R.-J.}\ \bibnamefont {Slager}},\ }\bibfield  {title}
  {\bibinfo {title} {Topological classification of crystalline insulators
  through band structure combinatorics},\ }\href
  {https://doi.org/10.1103/PhysRevX.7.041069} {\bibfield  {journal} {\bibinfo
  {journal} {Phys. Rev. X}\ }\textbf {\bibinfo {volume} {7}},\ \bibinfo {pages}
  {041069} (\bibinfo {year} {2017})}\BibitemShut {NoStop}%
\bibitem [{\citenamefont {Po}(2020)}]{Po2020}%
  \BibitemOpen
  \bibfield  {author} {\bibinfo {author} {\bibfnamefont {H.~C.}\ \bibnamefont
  {Po}},\ }\bibfield  {title} {\bibinfo {title} {Symmetry indicators of band
  topology},\ }\href {https://doi.org/10.1088/1361-648X/ab7adb} {\bibfield
  {journal} {\bibinfo  {journal} {Journal of Physics: Condensed Matter}\
  }\textbf {\bibinfo {volume} {32}},\ \bibinfo {pages} {263001} (\bibinfo
  {year} {2020})}\BibitemShut {NoStop}%
\bibitem [{\citenamefont {Freed}\ and\ \citenamefont
  {Moore}(2013)}]{Freed2013}%
  \BibitemOpen
  \bibfield  {author} {\bibinfo {author} {\bibfnamefont {D.~S.}\ \bibnamefont
  {Freed}}\ and\ \bibinfo {author} {\bibfnamefont {G.~W.}\ \bibnamefont
  {Moore}},\ }\bibfield  {title} {\bibinfo {title} {Twisted equivariant
  matter},\ }\href {https://doi.org/10.1007/s00023-013-0236-x} {\bibfield
  {journal} {\bibinfo  {journal} {Annales Henri Poincar{\'e}}\ }\textbf
  {\bibinfo {volume} {14}},\ \bibinfo {pages} {1927} (\bibinfo {year}
  {2013})}\BibitemShut {NoStop}%
\bibitem [{\citenamefont {Hughes}\ \emph {et~al.}(2011)\citenamefont {Hughes},
  \citenamefont {Prodan},\ and\ \citenamefont {Bernevig}}]{Hughes2011}%
  \BibitemOpen
  \bibfield  {author} {\bibinfo {author} {\bibfnamefont {T.~L.}\ \bibnamefont
  {Hughes}}, \bibinfo {author} {\bibfnamefont {E.}~\bibnamefont {Prodan}},\
  and\ \bibinfo {author} {\bibfnamefont {B.~A.}\ \bibnamefont {Bernevig}},\
  }\bibfield  {title} {\bibinfo {title} {Inversion-symmetric topological
  insulators},\ }\href {https://doi.org/10.1103/PhysRevB.83.245132} {\bibfield
  {journal} {\bibinfo  {journal} {Phys. Rev. B}\ }\textbf {\bibinfo {volume}
  {83}},\ \bibinfo {pages} {245132} (\bibinfo {year} {2011})}\BibitemShut
  {NoStop}%
\bibitem [{\citenamefont {Turner}\ \emph {et~al.}(2012)\citenamefont {Turner},
  \citenamefont {Zhang}, \citenamefont {Mong},\ and\ \citenamefont
  {Vishwanath}}]{Turner2012}%
  \BibitemOpen
  \bibfield  {author} {\bibinfo {author} {\bibfnamefont {A.~M.}\ \bibnamefont
  {Turner}}, \bibinfo {author} {\bibfnamefont {Y.}~\bibnamefont {Zhang}},
  \bibinfo {author} {\bibfnamefont {R.~S.~K.}\ \bibnamefont {Mong}},\ and\
  \bibinfo {author} {\bibfnamefont {A.}~\bibnamefont {Vishwanath}},\ }\bibfield
   {title} {\bibinfo {title} {Quantized response and topology of magnetic
  insulators with inversion symmetry},\ }\href
  {https://doi.org/10.1103/PhysRevB.85.165120} {\bibfield  {journal} {\bibinfo
  {journal} {Phys. Rev. B}\ }\textbf {\bibinfo {volume} {85}},\ \bibinfo
  {pages} {165120} (\bibinfo {year} {2012})}\BibitemShut {NoStop}%
\bibitem [{\citenamefont {Fu}\ and\ \citenamefont {Kane}(2007)}]{Fu2007}%
  \BibitemOpen
  \bibfield  {author} {\bibinfo {author} {\bibfnamefont {L.}~\bibnamefont
  {Fu}}\ and\ \bibinfo {author} {\bibfnamefont {C.~L.}\ \bibnamefont {Kane}},\
  }\bibfield  {title} {\bibinfo {title} {Topological insulators with inversion
  symmetry},\ }\href {https://doi.org/10.1103/PhysRevB.76.045302} {\bibfield
  {journal} {\bibinfo  {journal} {Phys. Rev. B}\ }\textbf {\bibinfo {volume}
  {76}},\ \bibinfo {pages} {045302} (\bibinfo {year} {2007})}\BibitemShut
  {NoStop}%
\bibitem [{\citenamefont {Fukui}\ \emph {et~al.}(2005)\citenamefont {Fukui},
  \citenamefont {Hatsugai},\ and\ \citenamefont {Suzuki}}]{Fukui2005}%
  \BibitemOpen
  \bibfield  {author} {\bibinfo {author} {\bibfnamefont {T.}~\bibnamefont
  {Fukui}}, \bibinfo {author} {\bibfnamefont {Y.}~\bibnamefont {Hatsugai}},\
  and\ \bibinfo {author} {\bibfnamefont {H.}~\bibnamefont {Suzuki}},\
  }\bibfield  {title} {\bibinfo {title} {{C}hern numbers in discretized
  brillouin zone: Efficient method of computing (spin) {H}all conductances},\
  }\href {https://doi.org/10.1143/JPSJ.74.1674} {\bibfield  {journal} {\bibinfo
   {journal} {Journal of the Physical Society of Japan}\ }\textbf {\bibinfo
  {volume} {74}},\ \bibinfo {pages} {1674} (\bibinfo {year} {2005})},\ \Eprint
  {https://arxiv.org/abs/https://doi.org/10.1143/JPSJ.74.1674}
  {https://doi.org/10.1143/JPSJ.74.1674} \BibitemShut {NoStop}%
\bibitem [{\citenamefont {Kartsev}\ \emph {et~al.}(2020)\citenamefont
  {Kartsev}, \citenamefont {Augustin}, \citenamefont {Evans}, \citenamefont
  {Novoselov},\ and\ \citenamefont {Santos}}]{Kartsev2020}%
  \BibitemOpen
  \bibfield  {author} {\bibinfo {author} {\bibfnamefont {A.}~\bibnamefont
  {Kartsev}}, \bibinfo {author} {\bibfnamefont {M.}~\bibnamefont {Augustin}},
  \bibinfo {author} {\bibfnamefont {R.~F.~L.}\ \bibnamefont {Evans}}, \bibinfo
  {author} {\bibfnamefont {K.~S.}\ \bibnamefont {Novoselov}},\ and\ \bibinfo
  {author} {\bibfnamefont {E.~J.~G.}\ \bibnamefont {Santos}},\ }\bibfield
  {title} {\bibinfo {title} {Biquadratic exchange interactions in
  two-dimensional magnets},\ }\href
  {https://doi.org/10.1038/s41524-020-00416-1} {\bibfield  {journal} {\bibinfo
  {journal} {npj Computational Materials}\ }\textbf {\bibinfo {volume} {6}},\
  \bibinfo {pages} {150} (\bibinfo {year} {2020})}\BibitemShut {NoStop}%
\bibitem [{\citenamefont {Basov}\ \emph {et~al.}(2017)\citenamefont {Basov},
  \citenamefont {Averitt},\ and\ \citenamefont {Hsieh}}]{Basov2017}%
  \BibitemOpen
  \bibfield  {author} {\bibinfo {author} {\bibfnamefont {D.~N.}\ \bibnamefont
  {Basov}}, \bibinfo {author} {\bibfnamefont {R.~D.}\ \bibnamefont {Averitt}},\
  and\ \bibinfo {author} {\bibfnamefont {D.}~\bibnamefont {Hsieh}},\ }\bibfield
   {title} {\bibinfo {title} {Towards properties on demand in quantum
  materials},\ }\href {https://doi.org/10.1038/nmat5017} {\bibfield  {journal}
  {\bibinfo  {journal} {Nature Materials}\ }\textbf {\bibinfo {volume} {16}},\
  \bibinfo {pages} {1077} (\bibinfo {year} {2017})}\BibitemShut {NoStop}%
\bibitem [{\citenamefont {Oka}\ and\ \citenamefont
  {Kitamura}(2019)}]{Takashi2019}%
  \BibitemOpen
  \bibfield  {author} {\bibinfo {author} {\bibfnamefont {T.}~\bibnamefont
  {Oka}}\ and\ \bibinfo {author} {\bibfnamefont {S.}~\bibnamefont {Kitamura}},\
  }\bibfield  {title} {\bibinfo {title} {Floquet engineering of quantum
  materials},\ }\href
  {https://doi.org/10.1146/annurev-conmatphys-031218-013423} {\bibfield
  {journal} {\bibinfo  {journal} {Annual Review of Condensed Matter Physics}\
  }\textbf {\bibinfo {volume} {10}},\ \bibinfo {pages} {387} (\bibinfo {year}
  {2019})},\ \Eprint
  {https://arxiv.org/abs/https://doi.org/10.1146/annurev-conmatphys-031218-013423}
  {https://doi.org/10.1146/annurev-conmatphys-031218-013423} \BibitemShut
  {NoStop}%
\bibitem [{\citenamefont {McIver}\ \emph {et~al.}(2020)\citenamefont {McIver},
  \citenamefont {Schulte}, \citenamefont {Stein}, \citenamefont {Matsuyama},
  \citenamefont {Jotzu}, \citenamefont {Meier},\ and\ \citenamefont
  {Cavalleri}}]{McIver2020}%
  \BibitemOpen
  \bibfield  {author} {\bibinfo {author} {\bibfnamefont {J.~W.}\ \bibnamefont
  {McIver}}, \bibinfo {author} {\bibfnamefont {B.}~\bibnamefont {Schulte}},
  \bibinfo {author} {\bibfnamefont {F.-U.}\ \bibnamefont {Stein}}, \bibinfo
  {author} {\bibfnamefont {T.}~\bibnamefont {Matsuyama}}, \bibinfo {author}
  {\bibfnamefont {G.}~\bibnamefont {Jotzu}}, \bibinfo {author} {\bibfnamefont
  {G.}~\bibnamefont {Meier}},\ and\ \bibinfo {author} {\bibfnamefont
  {A.}~\bibnamefont {Cavalleri}},\ }\bibfield  {title} {\bibinfo {title}
  {Light-induced anomalous {H}all effect in graphene},\ }\href
  {https://doi.org/10.1038/s41567-019-0698-y} {\bibfield  {journal} {\bibinfo
  {journal} {Nature Physics}\ }\textbf {\bibinfo {volume} {16}},\ \bibinfo
  {pages} {38} (\bibinfo {year} {2020})}\BibitemShut {NoStop}%
\bibitem [{\citenamefont {de~la Torre}\ \emph {et~al.}(2021)\citenamefont
  {de~la Torre}, \citenamefont {Kennes}, \citenamefont {Claassen},
  \citenamefont {Gerber}, \citenamefont {McIver},\ and\ \citenamefont
  {Sentef}}]{Torre2021}%
  \BibitemOpen
  \bibfield  {author} {\bibinfo {author} {\bibfnamefont {A.}~\bibnamefont
  {de~la Torre}}, \bibinfo {author} {\bibfnamefont {D.~M.}\ \bibnamefont
  {Kennes}}, \bibinfo {author} {\bibfnamefont {M.}~\bibnamefont {Claassen}},
  \bibinfo {author} {\bibfnamefont {S.}~\bibnamefont {Gerber}}, \bibinfo
  {author} {\bibfnamefont {J.~W.}\ \bibnamefont {McIver}},\ and\ \bibinfo
  {author} {\bibfnamefont {M.~A.}\ \bibnamefont {Sentef}},\ }\bibfield  {title}
  {\bibinfo {title} {Colloquium: Nonthermal pathways to ultrafast control in
  quantum materials},\ }\href {https://doi.org/10.1103/RevModPhys.93.041002}
  {\bibfield  {journal} {\bibinfo  {journal} {Rev. Mod. Phys.}\ }\textbf
  {\bibinfo {volume} {93}},\ \bibinfo {pages} {041002} (\bibinfo {year}
  {2021})}\BibitemShut {NoStop}%
\bibitem [{\citenamefont {Ron}\ \emph {et~al.}(2020)\citenamefont {Ron},
  \citenamefont {Chaudhary}, \citenamefont {Zhang}, \citenamefont {Ning},
  \citenamefont {Zoghlin}, \citenamefont {Wilson}, \citenamefont {Averitt},
  \citenamefont {Refael},\ and\ \citenamefont {Hsieh}}]{Ron2020}%
  \BibitemOpen
  \bibfield  {author} {\bibinfo {author} {\bibfnamefont {A.}~\bibnamefont
  {Ron}}, \bibinfo {author} {\bibfnamefont {S.}~\bibnamefont {Chaudhary}},
  \bibinfo {author} {\bibfnamefont {G.}~\bibnamefont {Zhang}}, \bibinfo
  {author} {\bibfnamefont {H.}~\bibnamefont {Ning}}, \bibinfo {author}
  {\bibfnamefont {E.}~\bibnamefont {Zoghlin}}, \bibinfo {author} {\bibfnamefont
  {S.~D.}\ \bibnamefont {Wilson}}, \bibinfo {author} {\bibfnamefont {R.~D.}\
  \bibnamefont {Averitt}}, \bibinfo {author} {\bibfnamefont {G.}~\bibnamefont
  {Refael}},\ and\ \bibinfo {author} {\bibfnamefont {D.}~\bibnamefont
  {Hsieh}},\ }\bibfield  {title} {\bibinfo {title} {Ultrafast enhancement of
  ferromagnetic spin exchange induced by ligand-to-metal charge transfer},\
  }\href {https://doi.org/10.1103/PhysRevLett.125.197203} {\bibfield  {journal}
  {\bibinfo  {journal} {Phys. Rev. Lett.}\ }\textbf {\bibinfo {volume} {125}},\
  \bibinfo {pages} {197203} (\bibinfo {year} {2020})}\BibitemShut {NoStop}%
\bibitem [{\citenamefont {Sriram}\ and\ \citenamefont
  {Claassen}(2022)}]{Sriram2022}%
  \BibitemOpen
  \bibfield  {author} {\bibinfo {author} {\bibfnamefont {A.}~\bibnamefont
  {Sriram}}\ and\ \bibinfo {author} {\bibfnamefont {M.}~\bibnamefont
  {Claassen}},\ }\bibfield  {title} {\bibinfo {title} {Light-induced control of
  magnetic phases in {K}itaev quantum magnets},\ }\href
  {https://doi.org/10.1103/PhysRevResearch.4.L032036} {\bibfield  {journal}
  {\bibinfo  {journal} {Phys. Rev. Res.}\ }\textbf {\bibinfo {volume} {4}},\
  \bibinfo {pages} {L032036} (\bibinfo {year} {2022})}\BibitemShut {NoStop}%
\bibitem [{\citenamefont {Monkhorst}\ and\ \citenamefont
  {Pack}(1976)}]{Monkhorst1976}%
  \BibitemOpen
  \bibfield  {author} {\bibinfo {author} {\bibfnamefont {H.~J.}\ \bibnamefont
  {Monkhorst}}\ and\ \bibinfo {author} {\bibfnamefont {J.~D.}\ \bibnamefont
  {Pack}},\ }\bibfield  {title} {\bibinfo {title} {Special points for
  brillouin-zone integrations},\ }\href
  {https://doi.org/10.1103/PhysRevB.13.5188} {\bibfield  {journal} {\bibinfo
  {journal} {Phys. Rev. B}\ }\textbf {\bibinfo {volume} {13}},\ \bibinfo
  {pages} {5188} (\bibinfo {year} {1976})}\BibitemShut {NoStop}%
\bibitem [{\citenamefont {Susner}\ \emph {et~al.}(2017)\citenamefont {Susner},
  \citenamefont {Chyasnavichyus}, \citenamefont {McGuire}, \citenamefont
  {Ganesh},\ and\ \citenamefont {Maksymovych}}]{Susner2017}%
  \BibitemOpen
  \bibfield  {author} {\bibinfo {author} {\bibfnamefont {M.~A.}\ \bibnamefont
  {Susner}}, \bibinfo {author} {\bibfnamefont {M.}~\bibnamefont
  {Chyasnavichyus}}, \bibinfo {author} {\bibfnamefont {M.~A.}\ \bibnamefont
  {McGuire}}, \bibinfo {author} {\bibfnamefont {P.}~\bibnamefont {Ganesh}},\
  and\ \bibinfo {author} {\bibfnamefont {P.}~\bibnamefont {Maksymovych}},\
  }\bibfield  {title} {\bibinfo {title} {Metal thio- and selenophosphates as
  multifunctional van der waals layered materials},\ }\href
  {https://doi.org/10.1002/adma.201602852} {\bibfield  {journal} {\bibinfo
  {journal} {Advanced Materials}\ }\textbf {\bibinfo {volume} {29}},\ \bibinfo
  {pages} {1602852} (\bibinfo {year} {2017})}\BibitemShut {NoStop}%
\bibitem [{\citenamefont {Gu}\ \emph {et~al.}(2019)\citenamefont {Gu},
  \citenamefont {Zhang}, \citenamefont {Le}, \citenamefont {Li}, \citenamefont
  {Xiang},\ and\ \citenamefont {Hu}}]{Gu2019}%
  \BibitemOpen
  \bibfield  {author} {\bibinfo {author} {\bibfnamefont {Y.}~\bibnamefont
  {Gu}}, \bibinfo {author} {\bibfnamefont {Q.}~\bibnamefont {Zhang}}, \bibinfo
  {author} {\bibfnamefont {C.}~\bibnamefont {Le}}, \bibinfo {author}
  {\bibfnamefont {Y.}~\bibnamefont {Li}}, \bibinfo {author} {\bibfnamefont
  {T.}~\bibnamefont {Xiang}},\ and\ \bibinfo {author} {\bibfnamefont
  {J.}~\bibnamefont {Hu}},\ }\bibfield  {title} {\bibinfo {title} {{Ni}-based
  transition metal trichalcogenide monolayer: A strongly correlated
  quadruple-layer graphene},\ }\href
  {https://doi.org/10.1103/PhysRevB.100.165405} {\bibfield  {journal} {\bibinfo
   {journal} {Phys. Rev. B}\ }\textbf {\bibinfo {volume} {100}},\ \bibinfo
  {pages} {165405} (\bibinfo {year} {2019})}\BibitemShut {NoStop}%
\end{thebibliography}%

\end{document}